\documentclass[aps,prd,twocolumn,showpacs,nofootinbib,showkeys,superscriptaddress,preprintnumbers]{revtex4-1}

\usepackage{graphicx,floatflt}
\usepackage{amsmath}
\usepackage{epsfig}

\newcommand{\dd}{\mathrm{d}} 
\newcommand{\Mpl}{M_\mathrm{pl}} 
\newcommand{\mpl}{m_\mathrm{pl}} 

\newcommand{\rr}{\mathrm}
\newcommand{\ns}{n_{\rr s} }
\newcommand{\fnl}{f_{\rr{NL}}}
\newcommand{\phic}{\phi_{\rr c}}

\newcommand{\Nend}{N_{\rr {end}}}
\newcommand{\Nexit}{N_{\rr {exit}}}

\newcommand{\xiend}{\xi_{\mathrm{end}}}
\newcommand{\chiend}{\chi_{\mathrm{end}}}
\newcommand{\fNL}{f_{\mathrm{NL}}}
\newcommand{\Ni}{N_{,i}}
\newcommand{\Nij}{N_{,ij}}

\newcommand{\Mpcinv}{\ \rr{Mpc}^{-1}}

\newcommand{\be}{ \begin{equation}}
\newcommand{\ee}{\end{equation} }
\newcommand{\ba}{ \begin{eqnarray}}
\newcommand{\ea}{\end{eqnarray} }

\begin{document}

\leftline{}
\rightline{TUM-HEP-885/13}

\title{Non-gaussianities and curvature perturbations from hybrid inflation}

\author{S\'ebastien Clesse} \email{s.clesse@tum.de}
\affiliation{Physik Department T70, James-Franck-Strasse,
Technische Universit\"at M\"unchen, 85748 Garching, Germany}
\author{Bj\"orn Garbrecht} \email{garbrecht@tum.de}
\affiliation{Physik Department T70, James-Franck-Strasse,
Technische Universit\"at M\"unchen, 85748 Garching, Germany}
\author{Yi Zhu} \email{yi.zhu@tum.de}
\affiliation{Physik Department T70, James-Franck-Strasse,
Technische Universit\"at M\"unchen, 85748 Garching, Germany}
\date{\today}

\begin{abstract}
For the original hybrid inflation as well as the supersymmetric
F-term and D-term hybrid models, we calculate the level of
non-gaussianities and the power spectrum of curvature
perturbations generated during the waterfall, taking into account the
contribution of entropic modes. We focus on the regime of mild
waterfall, in which inflation continues for more than about 60
e-folds $N$ during the waterfall.  We find that the associated $\fnl$
parameter goes typically from $ \fnl  \simeq -1 / \Nexit $ in the
regime with $N \gg 60 $, where $\Nexit$ is the number of e-folds between 
the time of Hubble exit of a pivot scale and the end of inflation, 
down to $\fnl \sim -0.3$ when $N \gtrsim 60 $,
i.e. much smaller in magnitude than the current bound from Planck.  Considering
only the adiabatic perturbations, the power spectrum is red, with
a spectral index $n_s = 1 - 4 / \Nexit $ in the case $N \gg 60 $,
whereas in the case $N \gtrsim 60 $ it increases up to unity.
Including the contribution of entropic modes does not change
observable predictions in the first case and the spectral index is too low for this regime to be viable.
In the second case, entropic modes are a relevant source for the power spectrum of curvature perturbations,
of which the amplitude increases by several orders of magnitudes. 
When spectral index values are consistent with observational constraints, the primordial spectrum amplitude is much larger than the observed value, and can
even lead to black hole formation. We conclude that due to the important contribution of entropic modes, the parameter space
leading to a mild waterfall phase is excluded by CMB observations
for all the considered models.
\end{abstract}
\maketitle

\section{Introduction}

In the standard cosmological scenario, the large scale structures
of the Universe are seeded by the quantum fluctuations of one or
more scalar fields during a primordial phase of exponentially
accelerated expansion.   Due to this phase of
inflation,  the quantum field fluctuations becomes classical and
are stretched
outside the Hubble radius.  They lead to a nearly scale invariant
power spectrum of curvature perturbations ${\cal P}_\zeta$, whose
amplitude and spectral index are constrained by CMB observations.
The recent results from the Planck experiment \cite{Ade:2013lta} give the 
best bounds with 
$\ln 10^{10} A_{\zeta} = 3.089 \pm 0.027$ and $ \ns = 0.9603 \pm 0.0073$ with $68\%$ C.L.,
where $A_\zeta$ is the value of ${\cal P}_\zeta(k_*)$ at the scale
$k_*=0.05 {\rm Mpc}^{-1}$.
Models of inflation predicting $\ns > 1$ are now ruled out at 
more than the $5 \sigma$ level \cite{Ade:2013rta}.

This is commonly thought to be the case of original
hybrid inflation \cite{Linde:1993cn}.   In this two-field model, inflation is
realized in the false vacuum along a nearly flat valley of the
field potential.  It ends with a nearly instantaneous waterfall
phase, triggered when the auxiliary field develops a tachyonic
instability.  Hybrid models are well motivated from the point of
view of particle physics because they can be embedded in
frameworks like supersymmetry~\cite{Dvali:1994ms, Copeland:1994vg, Binetruy:1996xj, Kallosh:2003ux, Clauwens:2007wc,Garbrecht:2006az}, supergravity~\cite{Halyo:1996pp,Binetruy:2004hh} and Grand Unified
Theories~\cite{Jeannerot:1997is,Jeannerot:2000sv, Fukuyama:2008dv}.   
Other realizations of hybrid inflation in the context
of string theories~\cite{Koyama:2003yc,Dvali:1998pa,Berkooz:2004yc, Davis:2008sa, Brax:2006yq, Kachru:2003sx},
and extra-dimensional theories~\cite{Fairbairn:2003yx}
have also  been proposed.

In addition to a blue tilted power spectrum,  the original hybrid
model was thought to suffer from a problem of fine-tuning of
initial field values~\cite{Mendes:2000sq,Tetradis:1997kp}.   
Moreover, because a $Z_2$
symmetry is broken at the end of inflation,  it leads to the
formation of domain walls with dramatic consequences for
cosmology.

 However, recent developments have shown that there exists a regime in the
parameter space of the original hybrid model for which the final
waterfall phase is sufficiently mild for inflation to continue
for more than 60 e-folds $N$ after the fields have crossed the critical
instability point~\cite{Clesse:2010iz,Kodama:2011vs}.  
In this scenario, topological defects are
stretched outside the observable Universe, thus escaping observation, and the
power spectrum of adiabatic perturbations is red,
possibly in agreement with CMB observations.  Moreover, the initial field values leading to
inflation have been shown to occupy an important proportion of the
field space in a large part of the potential parameter space~\cite{Clesse:2009ur,Clesse:2008pf,Clesse:2009zd,Easther:2013bga}, thus
solving the fine-tuning problem of initial conditions.  Similar conclusions have been
obtained for the most well-known supersymmetric realizations of the hybrid model,
the so-called F-term and D-term models~\cite{Clesse:2012dw}.  
In the standard regime of a fast waterfall, the contribution of the cosmic strings formed at the critical instability
point must be considered for these models.  The F-term model was nevertheless
found to be in tension with WMAP data
whereas the D-term model was strongly disfavoured~\cite{Battye:2010hg,Battye:2006pk}.  
With the Planck results, the degeneracy between the
spectral index and the string tension, that tends to favor larger values
for the spectral index, is strongly reduced and both models appear to be ruled out in the standard regime~\cite{Ade:2013xla}.

The revival of the
original hybrid scenario would be nevertheless of short duration
if the mild waterfall scenario leads generically
to a large level of non-gaussianities, with a local $\fnl $ parameter
outside the recent Planck bound $\fnl = 2.7 \pm 5.8$ ($68\%$ C.L.)~\cite{Ade:2013ydc}, 
or if the power spectrum of
curvature perturbations is strongly affected by the
entropic modes along the waterfall trajectories.

The aim of this paper is to evaluate the power spectrum of curvature
perturbations and the contribution of entropic modes, as well as
the level of non-gaussianities produced in the mild waterfall
regime, for the original hybrid scenario as well as for
the supersymmetric F-term and D-term models.  Our calculation
therefore differs from 
Refs.~\cite{Lyth:2012yp, Bugaev:2011qt, Bugaev:2011wy, Gong:2010zf, Fonseca:2010nk, Abolhasani:2010kr, Alabidi:2006wa, Lyth:2010ch,Enqvist:2004bk, Barnaby:2006km, Barnaby:2006cq, Alabidi:2006hg} 
since they consider only
the scenario of a waterfall phase lasting no more than a few e-folds.
A similar calculation was performed in Ref.~\cite{Mulryne:2011ni} for the original model, and some values for the parameters
were found to lead to a large level of local $\fNL$, but the focus there was on
super-planckian field evolution in a regime where the potential is dominated by the separable terms.
A similar analysis is also performed in Ref.~\cite{Avgoustidis:2011em} but only for a single trajectory. 
In this paper, we use an unified parametrization of the 
potential for the three considered models (F-term, D-term and original hybrid inflation)
and restrict the analysis
to sub-planckian field values, such that supergravity corrections to the potential can be
neglected. In this regime, the potential is not separable.

We use the $\delta N$ formalism to calculate the local
 $\fnl $ parameter, both analytically and numerically, as well as the amplitude and spectral
index of the power spectrum of curvature perturbations.  
As a cross-check, we have also integrated numerically the linear multi-field perturbations and derived the exact power spectrum of curvature perturbations.  This latter method has been chosen to study the time evolution of the field perturbations and their respective contribution to the adiabatic and entropic modes during the waterfall \footnote{Notice that separable universe techniques can also be used to study the time evolution of both curvature and iso-curvature perturbations~\cite{Langlois:2008vk}.}.     

Denoting by $\Nexit$ the number of e-folds between the time of Hubble exit of a pivot scale
and the end of inflation, we find that the associated $\fnl$ parameter goes typically from
$\fnl \simeq -1 / \Nexit $,
when the waterfall lasts for $N \gg 60 $ e-folds, down to $\fnl
\sim -0.3$ when $N \gtrsim 60 $.  In all cases the magnitude of $\fNL$ does not exceed the 
bounds of Planck.  Considering only the adiabatic
perturbations, the power spectrum is red, with a spectral index $n_s
= 1 - 4 / \Nexit $, in the case $N \gg 60 $, whereas in the case $N
\gtrsim 60 $, it increases up to unity, which apparently suggests
the presence of a parametric region, where the spectral index is in accordance with
the observation by Planck~\cite{Kodama:2011vs,Clesse:2012dw}. When including the contribution of entropic modes,
we find that  the predictions
do not change in the first case ($N \gg 60 $).  However, in the second
case ($N\gtrsim 60 $), entropic modes are a sizeable source for the power spectrum of curvature perturbations
and the spectral index first takes lower values before eventually
increasing up to unity.  However, the amplitude of the power spectrum of curvature perturbations is enhanced by several orders of
magnitudes and can even reach the level of black hole formation.
We conclude that due to the important contribution of entropic
modes, the parameter space leading to a mild waterfall phase is
excluded by CMB observations for all the considered models.

This paper is organised as follows:  in Sec.~\ref{sec:MF_background} we give the
exact background multi-field dynamics.  Then are introduced the $\delta N$ formalism (Sec. \ref{sec:deltaN})
and the linear theory of multi-field perturbations (Sec.~\ref{sec:MF_perturbations}).
In Sec.~\ref{sec:models} the considered hybrid models are briefly described and the unified parametrization of the potential is given.
Sec.~\ref{sec:waterfall_SR} is dedicated to the slow-roll dynamics during the mild waterfall phase. In Secs.~\ref{sec:non_gauss}
and~\ref{sec:power_spectrum}, we evaluate respectively the level of non-gaussianities
and the power spectrum of curvature perturbations, and
 compare the analytical and numerical results.
In the conclusion (Sec.~\ref{sec:conclusion}), we discuss the impact of our results on the constraints of hybrid models
and envisage interesting perspectives to this work.

\section{Multi-field background dynamics} \label{sec:MF_background}

Assuming that the Universe was filled with $n$ nearly homogeneous
real scalar fields $\phi_{i=1,2...,n} $, the background dynamics
is given by the Friedmann-Lema\^itre equations
\be \label{eq:FLtc12field}
H^2 = \frac {1 }{3 \Mpl^2}  \left[ \frac 1 2 \sum_{i=1}^n  \dot
\phi _i ^2 + V(\phi_{i=1,...,n})  \right] ~,
\ee
\be
\frac{\ddot a }{a} = \frac {1}{3 \Mpl^2} \left[ - \sum_{i=1}^n
\dot \phi_i^2
 + V(\phi_{i=1,...,n}) \right]~,
\ee
as well as by $n$ coupled Klein-Gordon equations
\be \label{KGtc2fieldd}
\ddot \phi_i + 3 H \dot \phi_i + \frac {\partial V}{\partial
\phi_i} = 0~,
\ee
where $H$ is the Hubble expansion rate, $\Mpl \equiv m_{\rr{pl}} / \sqrt{8 \pi}$ is the reduced Planck mass,
 $V(\phi_{i=1,...,n})$ is the field potential,
and where a dot denotes the derivative with respect to the cosmic time $t$.
Then one can define $\sigma $, the so-called \textit{adiabatic
field}~\cite{Gordon:2000hv}, that describes the collective
evolution of all the fields along the classical trajectory, and
the velocity field
\be \label{eq:adiabfield}
\dot \sigma \equiv \sqrt{\sum_{i=1}^n \dot \phi_i^2}~.
\ee
The equation of motion of the adiabatic field is given by
\be \label{KGtcadiab}
\ddot \sigma + 3 H \dot \sigma + V_\sigma = 0 ~,
\ee
where
\be
V_\sigma \equiv \sum_{i=1}^n u_i \frac{\partial V}{\partial
\phi_i}~,
\ee
with $u_i$ being the components of a unit vector along the field
trajectory $u_i \equiv \dot \phi_i / \dot \sigma $.

\section{The $\delta N$ formalism} \label{sec:deltaN}

The $\delta N $ formalism, based on the separate universe
approximation (and other assumptions discussed e.g.~in Ref.~\cite{Sugiyama:2012tj}),
states that the curvature
perturbation $\zeta(x,t)$ on a spatial hypersurface of uniform energy density
is given by the difference between the number of
e-folds realized from an initially flat hypersurface $N(t,x)\equiv \ln [\tilde a(t)/a(t_{\rr i})]$, where $\tilde a(x,t)$ is the local scale factor,
and the unperturbed number of e-folds $N_0(t) \equiv \ln[a(t)/a(t_{\rr i})]$.
If we label respectively by $\rr i$ and $\rr f$
the initial and final hypersurface, one has
\be
\zeta = \delta N _{\rr i} ^{\rr f} \equiv N(t,x) - N_0(t)~.
\ee
Our initial hypersurface is chosen at the time $t_*$
corresponding to the Hubble exit of the observable pivot scale $k_* = 0.05 \Mpcinv$ 
(in the rest of the paper, star subscripts indicate quantities evaluated at $t_*$).
The final hypersurface must be of uniform energy density.  In this paper we are interested in 
the curvature perturbations at the end of the slow-roll regime, when one of the slow-roll parameters reaches unity.

The field perturbations are close to Gaussian and have a very small
amplitude, so that the observed curvature perturbations are given
to good accuracy by
\be \zeta  \simeq \sum_{i=1}^n  N_{,i}  \delta
\phi_i  + \frac 1 2 \sum_{i,j = 1} ^n N_{,ij} \delta \phi_i \delta
\phi_j~,
\ee
where we have used the notation
\be N_{,i} \equiv \frac{\partial \delta N ^{\rr
f}_{\rr i}}{\partial \phi_i ^{\rr i}}, N_{,ij} \equiv \frac{\partial ^2
\delta N^{\rr f}_{\rr i} }{\partial \phi_i ^{\rr i} \partial \phi_j ^{\rr i}}~.
\ee
The amplitude of the reduced bispectrum is given by~\cite{Lyth:2005fi}
\be \label{eq:deltaN_fNL}
\fnl = - \frac 5 6 \dfrac{\sum _{i,j}  N_{,i} N_{,j}  N_{,ij} }
{\left( \sum_i N_{,i}^2  \right) ^2}.
\ee
One can also calculate
the power spectrum amplitude and spectral tilt as
\be \label{eq:deltaN_As2}
{\cal P}_\zeta(k_*)  = \frac{H_* ^2 }{4 \pi^2 }
\sum_i N_{,i}^2\,,
\ee
\be \label{eq:deltaN_ns}
\ns - 1 = - 2 \epsilon_{1*} + \frac{2 \sum
_{ij}\dot \phi_{i*} N_{,j} N_{,ij}  }{H_* \sum_i N_{,i}^2 }
\,,
\ee
where $\epsilon_{1} = - \dot H / H^2 $ is the first slow-roll parameter. Practically, instead of a final surface of uniform density, we have chosen a final surface of constant field value, more precisely the value taken by the inflaton when the non-perturbed trajectory breaks the slow-roll approximation.  As explained later in Section~\ref{sec:non_gauss}, the e-fold differences between these two surfaces is negligible compared to the e-folds variations between trajectories reaching them. This approximation leads therefore to accurate predictions.  In addition, we have checked numerically for all the considered parameter sets that these two possible final surfaces give identical observable predictions.

We furthermore need to introduce the
number of e-folds $N^t$ in the sense of a
reparameterization of time, $dN^t=H dt$, implying that during de Sitter inflation,
where $a(t)=\exp(Ht)$,
$N^t=H t$. We use the subscript
$k$ to indicate its value when a given scale $k$ exits the horizon,
$*$ when this scale is the pivot scale
and the subscript `end' to indicate its value at a uniform density
surface at the end of inflation. As stated above, the
latter can be practically obtained by
checking for the violation of the slow-roll conditions, {\it cf.} also
the discussion in Section~\ref{subsec:numeric}. 
Finally we define the function 
\be
\label{N:Nt}
N\equiv \Nend^t-N^t\,.
\ee
giving the number of e-folds up to the end of inflation.

\section{The linear theory of multi-field perturbations} \label{sec:MF_perturbations}

\subsection{Perturbed equations}

The perturbed metric can be written as
\be
\dd s^2= a^2 \left[- (1+2\Phi) \dd \eta^2 + (1-2\Psi) \delta_{ij} \dd x^i \dd x^j \right]
\,,
\ee
where $\Phi$ and $\Psi$ are the Bardeen potentials and $\eta$ is the conformal time,
which is related to the cosmic time $t$ as $dt=a d\eta$.
In the longitudinal
gauge, the spatial non-diagonal Einstein equations perturbed at first
order lead to $\Phi = \Psi$, such that the $(0,0)$, $(0,i)$ and
$(i,i)$ equations read
 \ba \label{Eq:pert_multifield1}
& & - 3 \mathcal H (\Phi'+\mathcal H \Phi )
+ \nabla ^2 \Phi \nonumber \\ 
& = & \frac {4 \pi}{m_{\mathrm pl}^2} \sum_{i=1}^n \left( \phi_i' \delta \phi_i' - \phi_i'^2 \Phi +  a^2 \frac{\partial V}{\partial \phi_i} \delta \phi_i \right) ~, 
\\  \label{Eq:pert_multifield2}
& & \Phi' + \mathcal H \Phi  =  \frac{4\pi}{m_{\mathrm pl}^2 } \sum_{i=1}^n \phi_i' \delta \phi_i ~,
  \\
 & & \Phi'' + 3 \mathcal H \Phi'
+  \Phi \left( 2\mathcal H ' + \mathcal H ^2 \right) \nonumber \\
& = & \frac {4 \pi}{m_{\mathrm pl}^2}  \sum_{i=1}^n  \left( \phi_i'
\delta \phi_i ' - \phi_i'^2 \Phi - a^2 \frac {\partial V}{\partial
\phi_i} \delta \phi_i \right) ~, \label{Eq:pert_multifield3}
\ea
where $\mathcal H \equiv a'/a$ and $\delta \phi_i $ is the perturbation of the scalar field
$\phi_i$, and where a prime denotes the derivative with respect to
the conformal time. On the other hand, the $n$ perturbed
Klein-Gordon equations read
\ba
& & \delta \phi_i '' + 2 \mathcal H \delta \phi_i' - \nabla^2 \delta
\phi_i + \sum_{j=1}^n a^2 \delta \phi_j \dfrac
{\partial^2V}{\partial \phi_i \partial \phi_j} \nonumber \\
\label{Eq:KGpert_multifield} & = & 2 (\phi_i'' + 2
\mathcal H \phi_i' ) \Phi + 4 \phi_i'  \Phi' ~.
\ea
The field perturbations are coupled to each other through
the cross derivatives of the potential and the Bardeen potential.
By adding Eq.~(\ref{Eq:pert_multifield1}) to
Eq.~(\ref{Eq:pert_multifield3}), and by using
Eq.~(\ref{Eq:pert_multifield2}), one obtains the evolution
equation for the Bardeen potential,
\ba
& & \Phi'' + 6 \mathcal H \Phi' + ( 2 \mathcal H' + 4 \mathcal H^2 )
\Phi - \nabla^2 \Phi \nonumber
\\ & = & - \frac {8 \pi}{m_{\mathrm pl}^2} a ^2
\sum_{i=1}^n \frac{\partial V}{\partial \phi_i} \delta \phi_i~.
\label{Eq:Bardeen_multifield}
\ea
We want to obtain the curvature perturbation $\zeta$, defined as
\be\label{eq:defPhi}
\zeta \equiv \Phi - \frac{\mathcal H}{\mathcal H' - \mathcal H^2
} (\Phi' + \mathcal H \Phi )~.
\ee
The background dynamics implies that $ \mathcal H' - \mathcal
H^2 = - 4 \pi \sigma'^2 / \mpl^2 $.  By using
Eq.~(\ref{Eq:pert_multifield2}), the comoving curvature can thus
be rewritten
\be
\zeta = \Phi + \frac{ \mathcal H}{\sigma'^2 }    \sum_{i=1}^n
\phi_i' \delta \phi_i ~.
\ee
From the background and the perturbed Einstein equations, one
can show that $\zeta$ evolves according to~\cite{Ringeval:2007am}
\ba
\zeta' & = &  \frac{m_{\rm pl}^2}{4\pi}\frac{\mathcal H}{\sigma'^2} \nabla^2 \Phi - \frac {2 \mathcal H}{\sigma'^2} \left[ a^2  \sum_{i=1}^n \phi_i' \frac{\partial V}{\partial \phi_i}  \right. \nonumber \\
& & \left. - \frac{a^2}{\sigma'^2} \left( \sum_{i=1}^n \phi_i' \frac{\partial V}{\partial \phi_i} \right)
\left(  \sum_{i=1}^n  \phi_i' \delta \phi_i\right)  \right] \\
 & = &  \frac{m_{\rm pl}^2}{4\pi} \frac{\mathcal H}{\sigma'^2} \nabla^2 \Phi
 - \frac {2 \mathcal H}{\sigma'^2} \bot_{ij} a^2 \frac{\partial V}{\partial \phi_i} \delta \phi_j ~,
\ea
where the orthogonal projector $\bot_{ij} \equiv \rr{Id} - u_i
u_j$ has been introduced. For a single field model, the second
term vanishes and one recovers the one-field evolution of $\zeta$.
For the multi-field case, one sees that entropy perturbations
orthogonal to the field trajectory can be a source of curvature
perturbations, even after Hubble exit.

\subsection{Numerical integration}


For the numerical integration of the perturbations, we refer to
Ref.~\cite{Ringeval:2007am} and give here only the guidelines for
the calculation of the exact power spectrum of curvature
perturbations in a multi-field scenario.  It is convenient to use
the number of e-folds as the time variable.  Some equations are redundant, and one can for instance 
use the Bardeen potential expressed
directly in terms of the field perturbations $\delta \phi_i$ and their derivatives.
After expanding in Fourier modes, Eq.~(\ref{Eq:KGpert_multifield}) reads~\cite{Ringeval:2007am}
\ba
& & \frac{\dd^2 \delta \phi_i}{\dd {N^t}^2} + (3 - \epsilon_{\rr 1} ) \frac{\dd \delta \phi_i}{\dd N^t} \nonumber \\
& + & \sum_{j=1}^n \frac{1}{H^2} \frac{\partial^2 V}{\partial \phi_i \partial \phi_j}  \delta \phi_j + \frac{k^2}{a^2 H^2} \delta \phi_i \nonumber \\
& = & 4 \frac{\dd \Phi}{\dd N^t} \frac{\dd \phi_i}{\dd N^t} - \frac {2 \Phi}{H^2} \frac{\partial V}{ \partial \phi_i} ~.
\ea
Here, $\delta\phi_i=\delta\phi_i(k,\eta)$ and we may use $k=|\mathbf k|$ because
of isotropy.

The quantization of the field perturbations in the limit
$k \gg a H$ provides initial conditions for the $\delta \phi_i $.
For the field operator, we take
\begin{align}
\delta\phi_i(\eta,\mathbf x)=\int\frac{d^3 k}{(2\pi)^3}
\Big[
a_i(\mathbf k){\rm e}^{-i\mathbf k\cdot\mathbf x}\delta\phi_i(k,\eta)+{\rm h.c.}
\Big]\,,
\end{align}
where h.c.~stands for hermitian conjugation and
$[a_i(\mathbf k),a^\dagger_j(\mathbf k^\prime)]=(2\pi)^3\delta_{ij}\delta(\mathbf k-\mathbf k^\prime)$.
The normalized quantum modes are defined
by
\be
v_{i,k}(\eta) = a \delta \phi_i(k,\eta)\,.
\ee
Neglecting the mass terms,
they obey the equation $v_{i,k}'' + k^2 v_{i,k} = 0$, and in the regime  $k \gg a H$,
one has
\be \label{eq:condinit_multifield}
\lim_{k/aH \rightarrow + \infty} v_{k,i} (\eta) =  k  \mathrm e^{-ik(\eta-\eta_{\mathrm i}) }~.
\ee
In terms of the field perturbations, the initial conditions
(denoted by the subscript i.c.~to avoid confusion with previous notation)
therefore read, up to a phase factor,
\ba
\delta \phi_{i,\rr{i.c.}} & = &\frac{1}{\sqrt{2k}} \frac{1}{a_{\rr{i.c.}}}~, \\
\left[ \frac{\dd \delta \phi_i}{\dd N^t} \right]_{\rr{i.c.}} & = & -
\frac{1}{a_{\rr{i.c.}}\sqrt{2k}}   \left( 1
+ i \frac{k}{a_{\rr{i.c.} } H_{\rr{i.c.}}}   \right)~.
\ea
It is not convenient to integrate the perturbations from the onset
of inflation, since the total number of e-folds can be much
larger than $N_*$.  In order to avoid the time consuming numerical
integration of sub-Hubble modes behaving like plane waves, it is
convenient to start to integrate the perturbations later, when the
 condition
\be \label{eq:Cq}
\frac{k}{ \mathcal H(n_{\rr{i.c.}} )}  = C_k \gg 1
\ee
is satisfied, where $C_k$ is a constant characterizing the
decoupling limit. To summarize, the numerical integration of multi-field perturbations can be divided in four steps:
\begin{enumerate}
\item The background dynamics is integrated until the end of inflation, such that $N^t_{\rr{end}} $ and $ N^t_{\rr{end}}  - N^t_* $ are obtained.
\item The background dynamics is integrated again, until $N^t_{\rr{i.c.}} $ is reached.  Initial conditions for the perturbations are fixed at this time.
\item For each comoving mode $k$, the background and the perturbation dynamics are integrated simultaneously from $N^t_{\rr{i.c.}} $ to $ N^t_{\rr{end}} $.
\item Determination of the scalar power spectrum, ${\cal P}_\zeta(k) = 1/(2 \pi^2) \sum_i |\zeta_i |^2$, where $i=1...n$ stands for the $n$ independent initial conditions for the field perturbations $\delta \phi_i$, $\zeta_i$ being the induced contributions to the curvature perturbation $\zeta$.
\end{enumerate}

\section{Hybrid models}\label{sec:models}

\subsection{Original version}

The original hybrid model of inflation was first proposed by
Linde~\cite{Linde:1993cn}.  Its potential reads
\be \label{eq:potenhyb2dNEW}
V(\phi,\psi) = \Lambda \left[  \left( 1 - \frac{\psi^2}{M^2}
\right)^2 + \frac{\phi^2}{\mu^2} + \frac{ 2 \phi^2
\psi^2}{\phi_{\rr c}^2 M^2}\right] .
\ee
The field $\phi$ is the inflaton, $\psi$ is an auxiliary
transverse field and $M,\mu,\phi_{\rr c}$ are three
parameters of mass dimension. Inflation is assumed to be realized in the
false vacuum~\cite{Copeland:1994vg} along the valley
$\langle\psi\rangle=0$.  In the usual description, inflation ends
when the transverse field develops a Higgs-type tachyonic
instability soon after the inflaton reaches the critical value $
\phi_{\rr c}  $.  From this point, the classical system is assumed
to evolve quickly toward one of its true minima
$\langle\phi\rangle=0$, $\langle\psi\rangle=\pm M$, whereas in a
realistic scenario one expects the instability to trigger a
tachyonic preheating era~\cite{Kofman:1997yn,Garcia-Bellido:1997wm,Felder:2000hj,Felder:2001kt,Copeland:2002ku,Senoguz:2004vu}.

\subsection{Supersymmetric F-term model}

Supersymmetric $F$-term inflation has been proposed
in Refs.~\cite{Dvali:1994ms,Copeland:1994vg} and has
subsequently been discussed extensively in the literature.
The underlaying details for our present analysis are given
in Ref.~\cite{Clesse:2012dw}, and we only present the
main assumptions and results for the potentials here.
The superpotential is given by
\be W=\kappa \widehat
S(\widehat {\bar H} \widehat H-{m}^2)~,
\ee
where the superfield $\widehat S$ is a
gauge singlet and the superfields $\widehat H$ ($\widehat{\bar H}$) transform
in the (anti-)fundamental representation of ${\rm
SU}({\cal N})$. This implies a tree-level scalar potential
\be
V_0=\kappa^2\left(|\bar HH-{m}^2|^2+|S\bar H|^2 + |S H|^2\right)\,,
\ee
where now $S$, $H$ and $ \bar H$ are complex scalar fields.
The $D$-term gives a large mass to the field
combination $(1/\sqrt 2)(H-\bar H)$, such that
we do not need to consider this field direction for the
dynamics of fluctuations during inflation. The waterfall field
can be identified as $\psi=(1/\sqrt 2)(H+\bar H\rangle)$ and the inflaton field
as $\phi=\sqrt 2|S|$. In terms of these degrees of freedom, the
tree-level potential is
\ba
\label{V:Ftree} V_0(\phi,\psi) &=& \kappa^2{m}^4 \left[
\left(1-\frac{\psi^2}{4{m}^2}\right)^2 +\frac{\phi^2\psi^2}{4{m}^4}
\right] \\
&=& \frac{\kappa^2}{4}\phi_{\rm c}^4 \left[
\left(1-\frac{\psi^2}{2\phi_{\rm c}^2}\right)^2
+\frac{\phi^2\psi^2}{\phi_{\rm c}^4} \right]~. \ea
In the potential valley where $\phi\geq\phi_{\rm c}=\sqrt{2} {m}$
and $\langle \psi\rangle=0$, the potential energy $\Lambda=\kappa^2 {m}^4$
spontaneously breaks supersymmetry. At one-loop order, this leads to
the following correction to the potential:
\ba \label{V:rad}
V_1 &=& \frac{\kappa^4{\cal N}}{128 \pi^2} \left[
(\phi^2-\phi_{\rm
c}^2)^2\log\left(\kappa^2\frac{\phi^2-\phi_{\rm
c}^2}{2Q^2}\right) + (\phi^2+\phi_{\rm
c}^2)^2 \right. \notag\\
& \times & \left. \log\left(\kappa^2\frac{\phi^2+\phi_{\rm
c}^2}{2Q^2}\right)
-2\phi^4\log\left(\frac{\kappa^2\phi^2}{2Q^2}\right) \right]\,,
\ea
where $Q$ is a renormalization scale. Notice that the derivatives
of this potential with respect to $\phi$ that are phenomenologically
relevant for inflation are independent of $Q$.

We are concerned in this work with the dynamics near the
critical point where $\phi\approx\phi_{\rm c}$ and
$\psi\approx 0$. As discussed in Ref.~\cite{Clesse:2012dw},
 treating the one-loop potential to linear order is a
good approximation in that regime. Of relevance is
the first derivative of the one-loop potential, \be
\label{DV:rad} \frac{\partial V_1(\phi)}{\partial
\phi}\Bigg|_{\phi=\phi_{\rm c}} =\frac{\kappa^4{\cal
N}}{8\pi^2}\phi_{\rm c}^3\log 2\,. \ee
In order to realize a substantial amount of e-folds below the
critical point, the relation
$\kappa\ll \phi_{\rm c}^2/M_{\rm P}^2$
must hold~\cite{Clesse:2012dw}. It can then easily
be derived that the second derivatives of the one-loop
potential may be neglected and that they
do not yield a phenomenologically relevant contribution to the spectral
tilt of the observable power spectrum~\cite{Clesse:2012dw}.

\subsection{Supersymmectric D-term model}

Here, the superpotential is~\cite{Binetruy:1996xj,
Halyo:1996pp} \be W=\kappa\widehat S \widehat{\bar H}\widehat H~,
\ee
where $\widehat S$ is again a singlet and
$\widehat H$ and
$\widehat{\bar H}$ transform according to the one-dimensional representation of a
${\rm U}(1)$ gauge group.
Spontaneous symmetry breaking is induced by the $D$-term
\be D=\frac g2 \left(|H|^2-|\bar H|^2
+m_{\rm FI}^2\right)~, \ee
where $m_{\rm FI}$ is the Fayet-Iliopoulos
term. The inflaton field is again $\phi=\sqrt
2|S|$, whereas now the waterfall field is given by $\psi=\sqrt 2 |\bar H|$.
The critical point, where $\psi$ becomes tachyonically unstable is
here
\be \phi_{\rm
c}=\frac1{\sqrt{2}}\frac g\kappa m_{\rm FI}\,. \ee
In terms of these various fields and variables, the tree-level potential
can be expressed as
\begin{align}
\label{V:Dtree} V_0&=\kappa^2\left( |H\bar H|^2+|SH|^2+|S\bar H|^2
\right) +\frac12 D^2
\\\notag
&=\frac{g^2}{8}m_{\rm FI}^2 \left[ \left(1-\frac{\psi^2}{2m_{\rm
FI}^2}\right)^2+2\frac{\kappa^2}{g^2}\frac{\phi^2\psi^2}{m_{\rm
FI}^4} \right]
\\\notag
&=\frac{\kappa^4}{2 g^2}\phi_{\rm c}^4 \left[ \left(
1-\frac{g^2}{4\kappa^2\phi_{\rm c}^2}\psi^2 \right)^2
+\frac{g^2}{2\kappa^2\phi_{\rm c}^4}\phi^2\psi^2 \right]\,.
\end{align}
The one-loop potential is readily obtained from its expression in
the $F$-term case~(\ref{V:rad}) when setting ${\cal N}=1$.

\subsection{Unified parametrization}

In this paper, we study the original hybrid model as well as
the supersymmetric F-term
and D-term variants in a unified approach, by
considering the following parametrization of the two-field potential,
\be \label{eq:Vunified} V(\phi,\psi) = \Lambda \left[ \left( 1 - \frac{\psi^2}{M^2}
\right)^2  + \left( \frac{\phi}{\mu}  \right)^p +  \frac{2 \phi^2
\psi^2}{ M^2 \phi_c^2} \right]\,, \ee
where $M$, $\phic $ are respectively the position of the global
minima and of the critical point of instability along the valley.
In the case of the original model, one has $p = 2$ whereas the
dynamics for the F-term and D-term models near the instability
point is well described when setting $ p = 1$~\cite{Clesse:2012dw}.
The relations between the
parameters of this potential and the model parameters for F-term
and D-term inflation are given in TABLE~\ref{Table:pars}.

\begin{table}[ht!]
\begin{center}
\begin{tabular}{|c||c|c|}
\hline
 & $F$-term & $D$-term\\
\hline
$\Lambda$&$\kappa^2 {m}^4$&$\frac{\kappa^4}{2 g^2}\phi_{\rm c}^4=\frac{g^2}{8} m_{\rm FI}^4$\\
\hline
$\phic$&$\sqrt 2 {m} $&$\frac{g}{\sqrt 2\kappa} m_{\rm FI}$\\
\hline
$ M $&$ 2 {m} $&  $ \sqrt 2 m_{\rm FI} $\\
\hline
$1 / \mu $& $\frac{\sqrt 2 \mathcal N \kappa^2 \log(2)} { 4 \pi^2 {m} }$ & $\frac{\sqrt 2\kappa g \log 2}{ 4 \pi^2 m_{\rm FI}}$\\
\hline
\end{tabular}
\end{center}

\caption{\label{Table:pars} Parameters to be substituted into the
potential~(\ref{eq:Vunified}) in order to obtain the $F$- and
$D$-term models close to the critical point.} 
\end{table}

For the purpose of deriving the phenomenological consequences of
hybrid inflation,
it is useful to note the derivatives
\begin{subequations}
\begin{align}
\label{DVDphi} \frac{\partial V}{\partial \phi}=& \frac{p \Lambda
\phi^{p-1} }{\mu^p} \left( 1 + \frac{4 \mu^p \phi^{2-p} \psi^2}{p M^2 \phic^2}
\right) \ ,
\\
\label{DVDpsi} \frac{\partial V}{\partial \psi}=& \frac{4 \psi
\Lambda}{M^2}  \left(\frac{\phi^2 - \phic^2}{\phic^2} +
\frac{\psi^2}{M^2} \right)\ ,
\\
\frac{\partial^2 V}{\partial\phi^2} =& \frac{p (p-1) \Lambda
\phi^{p-2} }{\mu^p} + \frac{4 \Lambda \psi^2}{M^2 \phic^2} \,,
\\ \frac{\partial^2 V}{\partial\psi^2} =& \frac{4 \Lambda}{M^2} \left( \frac{\phi^2-\phic^2}{\phic^2} + \frac{3 \psi^2}{M^2}  \right) \,,
\\
\frac{\partial^2 V}{\partial\phi\partial\psi} =&  \frac{8 \Lambda
\psi \phi }{M^2 \phic^2}\,.
\end{align}
\end{subequations}
In particular,
the first derivatives enter the slow-roll equations of motion, \be
\label{eom:slowroll} 3H \dot \phi =-\frac{\partial V}{\partial
\phi}\,, \hspace{5mm} 3 H \dot \psi = -\frac{\partial V}{\partial
\psi}\,, \hspace{5mm}  H^2 = \frac{V}{3 \Mpl^2}~. \ee
We also make use of the standard definition for the slow-roll parameters
$\eta_{XY}=\Mpl^2[\partial ^2 V/(\partial X\partial Y)]/V$,
where $X$ and $Y$ can be either of the canonically normalized fields
$\phi$ and $\psi$.

\section{Inflation along waterfall trajectories}\label{sec:waterfall_SR}

We are interested in the field dynamics during the waterfall regime,
i.e. in the times after the field trajectories cross the critical instability point
$\phi_c$.   Tachyonic preheating is triggered when the
exponentially growing long-wavelength perturbations of the auxiliary field $\psi$
become non-linear.  This only occurs when the tachyonic auxiliary field mass is
larger than the Hubble expansion rate, $
 m_\psi ^2 > H^2 \simeq V /(3 \Mpl^2),
 $
or equivalently when $-\eta_{\psi \psi} \gtrsim 1$. In the commonly
studied large-coupling scenarios of hybrid inflation,
this condition is satisfied after no more than a few e-folds after entering
the waterfall phase.
However, there exists a large
region in the parameter space (the small coupling limit for the
F-term and D-term models) for which the waterfall phase lasts for
more than 60 e-folds. We refer to this situation as the mild waterfall. The field
evolution in this regime has been studied for the original hybrid
model in Refs.~\cite{Clesse:2010iz,Kodama:2011vs}, and for the F-term and D-term models in
Ref.~\cite{Clesse:2012dw}.
In this Section, we re-derive the mild waterfall dynamics for the unified potential of Eq.~(\ref{eq:Vunified}). For the purpose of an analytic description, we follow Ref.~\cite{Kodama:2011vs}
and divide the field evolution below the critical point
into three phases:
\begin{itemize}
\item Phase 0:  when the second term of Eq.~(\ref{DVDpsi}) and the first term of Eq.~(\ref{DVDphi}) dominate.
\item Phase 1:  when the first term of Eq.~(\ref{DVDpsi}) and the first term of Eq.~(\ref{DVDphi}) dominate.
\item Phase 2:  when the first term of Eq.~(\ref{DVDpsi}) and the second term of Eq.~(\ref{DVDphi}) dominate.
\end{itemize}
Classically, the phase 0 does not last more than about one e-fold~\cite{Kodama:2011vs}.  
Moreover, in a realistic scenario, the quantum diffusion of the
auxiliary field dominates over the classical dynamics very close to
the critical instability point. 
For these reasons we do
not consider the phase 0 but only the phases 1 and 2.

Following the derivation of Ref.~\cite{Kodama:2011vs}, it is convenient to parameterize the fields as
\be
\label{param:exp}
\phi \equiv \phic \rr e^{\xi}~, \hspace{10mm} \psi \equiv \psi_{0}
\rr e^{\chi}
\,,
\ee
where $\psi_{0}$ is the initial condition for the
auxiliary field at the critical point of instability. 
During the slow-roll waterfall regime, one has $\xi<0$ and $|\xi|\ll1$,
which is consistently verified by the explicit solutions. 

\subsection{Phase 1}
\label{sec:waterfall_SR:ph1}

Solving the slow-roll equations, one finds that the field trajectories
during the phase 1 follow the relation
\be\label{xi:chi:ph1} \xi^2  = \frac{p M^2
\phic^{p-2}}{4 \mu^p } (\chi - \chi_i) 
\ee 
as long as the temporal
minimum (corresponding to the ellipse where $\dd V / \dd \psi$ vanishes) is not reached. In this case, phase 1 connects to phase 2
at the point $(\xi_2, \chi_2)$, with \be
\label{chi:2}
 \chi_2  \equiv  \ln  \left( \frac{\sqrt p\phic^{\frac p2} M}{ 2  \mu^{p/2} \psi_{0}}  \right)~.
  \ee
One obtains the number of e-folds realized in phase 1 by
integrating
 \be 
\label{dxi:dN:ph1}
\frac{\dd \xi}{\dd N^t} \simeq - \frac{p \phic^{p-2}
\Mpl^2 }{ \mu^p} \,. \ee

Imposing that $N^t =0$ at the critical point,
one finds \be\label{xi:N:ph1} \xi = - \frac{p \Mpl^2 \phic^{p-2} }{\mu^p}
N^t\,.\ee 
The temporal minimum for the field $\psi$ is located on
the trajectories
\be \label{eq:TM}
\xi = - \frac{\psi_{0}^2 \rr e^{2 \chi} }{2 M^2}~.
\ee
It is reached during phase 1 if the condition
\be \chi_2 < \frac{p \phic ^{p+2}}{ 16 \mu^p M^2} \ee is satisfied. 
One therefore gets 
\be
\label{xi2:chi2}
\xi_2 \equiv \left\{
  \begin{array}{rcr}
    - \dfrac{\sqrt p \phic^{\frac p2 -1} M}{2 \mu^{p/2} }\sqrt \chi_2& \rr{for}  & \chi_2 > \dfrac{p \phic ^{p+2}}{16 \mu^p M^2} \\
    - \dfrac{ p \phic^p}{8 \mu^p} & \rr{for}  & \chi_2 < \dfrac{p \phic ^{p+2}}{16 \mu^p M^2}  \\
  \end{array}
\right.\,. \ee The number of e-folds $N_1$ realized in phase
1 then follows as
\be N_1= \left\{
  \begin{array}{rcr}
    \dfrac{\sqrt{\chi_2}  \mu^{p/2} M  }{2 \sqrt p \phic^{\frac p2 -1} \Mpl^2} & \rr{for}  & \chi_2 > \dfrac{p \phic ^{p+2}}{16 \mu^p M^2} \\
    \dfrac{\phic^2}{ 8 \Mpl^2 } & \rr{for}  & \chi_2 <  \dfrac{p \phic ^{p+2}}{16 \mu^p M^2}\\
  \end{array}
  \,.
\right.
\ee

\subsection{Phase 2}

The slow-roll equations in phase 2 yield the
trajectory
 \be \xi^2 = \xi_2 ^2 + \frac{p M^2 \phic^{p-2}}{8 \mu^p} \left[ \rr e
^{2(\chi - \chi_2)} - 1 \right]~. \ee
Once again, we have to distinguish between the possibility where this trajectory
reaches the temporal minimum at some point before the end of inflation, and the possibility
where the temporal minumum is not reached during inflation.
In phase 2, the temporal minimum is reached when
\be \xi = \xi_{2
\rr{T.M.}} \equiv - \frac{M^2 }{2 \phic^2} -\sqrt{\frac{M^4}{4\phic^4}+\xi_2^2-\frac{p\phic^p}{8\mu^p}}
\,.
\ee
On the other hand, inflation ends before reaching the temporal minimum
when
\be
\eta_{\psi \psi}  
\simeq \frac{8 \Mpl^2 \xi}{M^2}\simeq - 1~.
\ee
In the opposite case, where the temporal minimum is reached before the end
of inflation, the slow-roll conditions are violated when
\be \eta_{\phi \phi} \simeq \frac{4 \Mpl^2 \psi^2}{ \phic^2
M^2 }\simeq 1\,.\ee
By using Eq.~(\ref{eq:TM}), one thus finds that 
\be 
\label{xi:end}
\xi_{\rr{end}}= \left\{
  \begin{array}{rcr}
  - \dfrac{\phic^2}{8 \Mpl^2} & \rr{for}  &  |\xi_{\rr{end}}|>|\xi_{2{\rm T.M.}}|\\
- \dfrac{M^2}{8 \Mpl^2} & \rr{for}  &  |\xi_{\rr{end}}|<|\xi_{2{\rm T.M.}}| \\
  \end{array}
\right. \,.\ee
During phase 2 and before reaching the temporal minimum,
assuming that $\xi \ll 1$ and $\chi_2 > 1/2 $, the
slow-roll equations in e-fold time can be solved exactly and one
finds~\cite{Kodama:2011vs}
\be \xi (N^t) = c'\sqrt{\frac{p M^2}{2\phic^{p-2}\mu^p}} \frac{(c'-c) f(N^t) - c' -
c}{(c'-c) f(N^t) +c' +c}\,, \ee where $c \equiv \sqrt{\chi_2 / 2} $,
$c' \equiv \sqrt{c^2 - 1/4} $ and \be f(N^t) = \exp \left[ \frac{8\sqrt 2
c' p \phic^{\frac p2 -1}\Mpl^2 (N^t-N_1) }{\sqrt{p\mu^p M^2}} \right]~. \ee
A good approximation can be obtained by considering the limit
$|\xi| \gg |\xi_2| $, where one can obtain 
\be
\label{xi:N:ph2}
\frac1\xi-\frac1{\xi_{\rm end}}
=\frac{8 \Mpl^2}{M^2} (N^t-\Nend^t)~. 
\ee

The number of e-folds realised along the temporal minimum is obtained by integrating
 \be \frac{\dd \xi}{\dd N^t} = \frac{8 \Mpl^2 \xi
}{\phic^2}~. \ee 
This gives
\be
\label{N:TM}
\Nend^t - N^t_{2 \rr{T.M.}} = \frac{\phic^2}{8 \Mpl^2} \ln \left(
\frac{\xiend}{\xi_{2\rr{T.M.}}} \right)~, \ee
where $N^t_{2 \rr{T.M.}}$ is the value of the parameter
$N^t$ when $\xi=\xi_{2\rr{T.M.}}$.
If we understand the model specified by the potential~(\ref{eq:Vunified})
as an effective theory valid below the Planck scale, we should impose that
$\phic\ll\Mpl$ and consequently, the number of e-folds inflation
continues for after reaching
the temporal minimum is very low.

\section{Non gaussianities from the mild waterfall phase}\label{sec:non_gauss}

\subsection{Analytical results} 

In this Section, we calculate the level of
non-gaussianities using the $\delta N$ formalism.
We denote the time where the transition between the
phases~1 and~2 occurs by $t_{1,2}$ and the time of
Hubble exit of the pivot scale $k_*$ by $t_*$.
We need to distinguish the case when the pivot
scale leaves the Hubble radius in phase~2 (case 1: long waterfall
phase with $\Nend^t \gg 60$, 
and $t_*>t_{1,2}$) from the
case where horizon exit occurs in
phase~1 (case 2: moderately long waterfall phase, 
with $\Nend^t \gtrsim 60$ and $t_*<t_{1,2}$).

\subsubsection{Case 1:  Hubble exit in phase 2}

The first step for calculating the local $\fNL$ parameter with the $\delta N$ formalism
 is to derive the number of e-folds until the end of inflation
starting from an arbitrary initial point $(\xi_i,\chi_i)$ in field space.
In order to compare with the observational bounds, we then use this result
to evaluate the
quantities $N_{,\phi}$, $N_{,\psi}$, $N_{,\phi \phi}$, $N_{,\phi \psi}$ and
$N_{,\psi \psi}$ for the pivot scale $k_*$, at which the unperturbed background
fields take the values $(\xi_*,\chi_*)$.

We first consider the case where inflation ends before the temporal minimum is reached. 
Integrating the slow-roll equation of motion in phase 2 give the trajectories
\be \label{eq:psi2phase2} \psi^2 =\psi_0 ^2 e ^{2 \chi}= 2
\phic^2 (\xi^2 - \xi_i^2 ) + \psi_0 ^2 \rr e ^{2 \chi_i}
\,.
\ee
(This equation also determines $\chiend$ when replacing
$\xi\to\xiend$ and $\chi\to\chiend$.)
Above relation can be used in combination with the
slow-roll equation for $\phi$ to derive the number of e-folds
that elapse while the fields evolve from
$(\xi_i,\chi_i)$ to $(\xiend, \chiend)$,
\be
\label{n:ph2:integral}
\Nend^t - N_i^t =
- \frac{M^2}{8 \Mpl^2} \int_{\xi_i}^{\xiend} \frac{\dd \xi}{\xi^2
- \xi_i ^2 + \frac{\psi_0 \rr e^{2 \chi_i}}{2 \phic^2}}
\,.
\ee
Defining $C \equiv - \xi_i ^2 + \psi_0 \rr e^{2 \chi_i} / (2
\phic^2)$, one finds
\begin{align} 
\Nend^t - N^t_i =& - \frac{M^2}{8 \Mpl^2
\sqrt{C}}
\notag\\
\times&
 \left[ \arctan \left(\frac{\xiend}{\sqrt C} \right)  -
\arctan \left(\frac{\xi_i}{\sqrt C} \right)  \right]
\label{eq:Nend-Ni_phase2a}
\end{align}
if $C>0$, and
\begin{align}
\Nend^t - N^t_i =& - \frac{M^2}{8 \Mpl^2
\sqrt{C}}
\notag\\
\times&
\left[ \text{arctanh} \left(\frac{\xiend}{\sqrt C} \right)  -
\text{arctanh} \left(\frac{\xi_i}{\sqrt C} \right)  \right]
\label{eq:Nend-Ni_phase2b}
\end{align}
if $C<0$.

For applying the $\delta N$ formalism, we need to evaluate the e-fold derivatives with respect to $\phi_i$ and $\psi_i$, evaluated in 
the limit where $\xi_i \rightarrow \xi $ and $\chi_i \rightarrow \chi$, where $\xi$ and $\chi$ belong to the
unperturbed waterfall trajectory, and for which the relations
\be \psi_0 ^2 \rr e^{2
\chi} =  2 \phic^2 \xi^2 \ee
and
\be \xi = - \frac{M^2}{8 \Mpl^2 \left( \Nend^t-N^t - \frac{M^2}{8
\Mpl^2 \xiend} \right)}~.
\ee
are satisfied.  Note that in this limit, one gets $C\rightarrow 0$ and 
one can expand Eqs.~(\ref{eq:Nend-Ni_phase2a}) and (\ref{eq:Nend-Ni_phase2b}) 
in a Taylor series to obtain
\be \label{eq:Ninphase2}
\Nend^t - N_i^t = \frac{M^2}{8 \Mpl^2} \left(
\frac{1}{\xiend} - \frac{1}{\xi_i} + \frac{|C|}{3 \xi_i^3} -
 \frac{|C|}{3 \xiend^3}   \right)~,
\ee
which is a consistent generalization of Eq.~(\ref{xi:N:ph2}).
We use this expression to calculate the e-fold derivatives
 $N_{,\phi}, N_{,\psi}, N_{,\phi \phi}, N_{,\phi \psi}$ and $N_{,\psi \psi}$
around the field configuration $(\xi_i,\chi_i)$. For the purpose of
calculating these derivatives, we can now
relabel $(\xi_i,\chi_i)\to(\xi,\chi)$. [I.e. $\xi$ has now a different purpose
than the integration variable in Eq.~(\ref{n:ph2:integral})].
We thus obtain
\ba
N_{,\phi} & = & \frac{M^2}{8  \phic \rr e^{\xi} \Mpl^2}
\left( \frac{1}{\xi^2} - \frac{2}{3 \xi^2} + \frac{2 \xi}{3 \xiend^3} \right) \notag\\
 & \simeq & \frac{M^2}{24  \phic \Mpl^2 \xi^2}
\label{eq:Nphi_phase2}
\ea
and
\ba N_{,\psi} & = & \frac{M^2}{8
  \Mpl^2} \left( \frac{\psi}{3 \phic^2 \xi^3} - \frac{\psi}{3 \phic^2 \xiend^3} \right) \notag\\
& \simeq & \frac{M^2  \psi}{24  \phic^2  \Mpl^2 \xi^3}
\,. \label{eq:Npsi_phase2}
\ea
The approximations above are valid when
$|\xi| < |\xiend |$ and thus $|\xi^3| \ll |\xiend^3 |$.
As we eventually replace $\xi\to\xi_*$, these relations are well
satisfied for the present purposes.
By
using $\psi = - \sqrt 2 \phic \xi $, we notice the
useful relation
\be
N_{,\phi} = - \frac{1}{\sqrt {2}} N_{,\psi}\,. \ee

For the second derivatives, when keeping only the leading terms, we find 
\be
N_{,\psi \psi} =
\frac{M^2}{24 \phic^2 \Mpl^2} \left( \frac{1}{ \xi^3} -
\frac{1}{ \xiend^3} \right) \simeq \frac{M^2}{24 \phic^2 \Mpl^2
\xi^3}\,, \label{eq:Npsipsi_phase2} \ee
\be
N_{,\phi \phi} \simeq \frac{4
M^2}{24 \phic^2 \Mpl^2 \xi^3} \label{eq:Nphiphi_phase2}\ee 
and
\be 
N_{,\phi \psi} =   -\frac{ M^2 \psi_k}{8 \phic^3 \rr e^{\xi} \Mpl^2 \xi^3 }  
\simeq   \frac{ 3 \sqrt 2  M^2}{24 \Mpl^2 \phic^2 \xi^3}
\,. \label{eq:Nphipsi_phase2}
\ee
One can now evaluate the $\fNL$
parameter at the pivot scale $k_*$.  We first notice that
\be
N_{,\phi \phi} = 4 N_{,\psi \psi} \simeq \frac{4}{3 \sqrt 2} N_{,\phi \psi}\,.
\ee
Using that the number of e-folds inflation lasts after
the horizon exit of the pivot scale is given by
$\Nexit=\Nend^t-N^t_*$ and evaluating above derivatives
for $(\xi,\chi)=(\xi_*,\chi_*)$, we obtain
  \be  \label{eq:fNLphase2}
\fNL \simeq  \frac{5}{18} \left( \frac{24 \Mpl^2 \xi_*}{M^2} \right) \simeq - \frac{5}{3 (\Nexit -
\frac{M^2}{8 \Mpl^2 \xiend})} \ll 1
\,.        
\ee 
In the case of the F-term
and D-term models, this expression reduces to
\be \label{eq:fNL_phase2}
\fNL \simeq - \frac{5}{3
(\Nexit +1)}~. \ee
 The level of non-gaussianities
 is therefore negative and very low, typically $\fNL \approx -0.03$ for
the F-term and D-term models.

We finally consider the situation where the temporal minimum
for the field $\psi$ is reached in phase~2
before the end of inflation. Then, we need to
evaluate
\be
N = N_* ^{\rr{T.M.}} + N_{\rr{T.M.}}^{\rr{end}}\,.
\ee
The first term on the right hand side is the number of e-folds realised between $t_*$ and the time where the temporal
minimum is reached.
The second term is the number of e-folds realised along the temporal minimum up to the end of inflation.  

Starting from an arbitrary
initial field value in phase 2, one finds the value of $\xi$ at which the trajectory crosses the temporal minimum, 
\be \label{eq:TM_phase2_from_ic}
\xi_{2\rr{T.M.}} = - \frac{M^2 }{2 \phic^2} -\sqrt{\frac{M^4}{4\phic^4}+ \xi_i^2+\psi_i^2/(2 \phic^2)}
\ee 
For evaluating $N ^{\rr{T.M.}}_{*,\phi}$, one can take Eq.~(\ref{eq:Ninphase2}) and replace $ \xiend $ by $\xi_{2\rr{T.M.}}$. For the derivatives, we again replace
$(\xi_i,\chi_i)\to(\xi,\chi)$ and
notice that 
\be
\frac{\dd\xi_{2\rr{T.M.}}}{\dd \xi} =\frac{-\xi}{\frac{M^2}{2 \phic^2} - \xi_{2\rr{T.M.}} }
\,.
\ee
This gives an additional term $- (\dd\xi_{2\rr{T.M.}} / \dd \phi_i) / \xi_{\rr{T.M.}}^2 $ in the
round brackets of Eq.~(\ref{eq:Nphi_phase2}), 
that is negligible compared to the leading term.
Next, using Eq.~(\ref{N:TM}), one can calculate 
\be
 N ^{\rr{end}}_{\rr{T.M.},\phi} = \frac{\phic}{8 \Mpl^2  \xi_{2\rr{T.M.}}}  \frac{\dd\xi_{2\rr{T.M.}}}{\dd \xi}=\frac{\phic}{\Mpl^2}{\cal O}(\xi)
\,.
\ee
Since $|\xi|<|\xiend|=\phic^2/(8\Mpl^2)$, we find
that $ N ^{\rr{end}}_{\rr{T.M.},\phi} \ll N ^{\rr{T.M.}}_{*,\phi}$,  
provided $\phic\ll\Mpl$, as one should require for the effective theory description
to be valid.

In a similar way, one finds that the leading terms in $N^{\rm end}_{{\rr T.M.},\psi}$, $N^{\rm end}_{{\rr T.M.},\psi \psi}$, $N^{\rm end}_{{\rr T.M.},\phi \psi}$ and $N^{\rm end}_{{\rr T.M.}, \phi,\phi}$, as well as in $\fNL$,
 are not modified when the temporal minimum is reached, apart that one has to replace $\xiend$ by $\xi_{\rr{T.M.}}$. 
As a consequence, the level of non-gaussianities is reduced
compared to the contributions from $N_*^{{\rr T.M.}}$
 and cannot in any case increase up to an observable level.

\subsubsection{Case 2:  Hubble exit in phase 1}

Now we consider the situation where the pivot scale exits the Hubble radius in the phase~1.  We first consider the case where the temporal minimum is not reached.
In order to obtain the spectra using the $\delta N$ approach, we need to generalize the
analysis of Section~\ref{sec:waterfall_SR:ph1}, where we have fixed the initial value of
$\xi$ to zero, to more general initial values $\xi_i$, such that we can obtain the necessary
e-fold derivatives.

By integrating the slow-roll equations of motion, we obtain the trajectory
\be
\xi^2 -  \xi_i^2 = \frac{p \phic^{p-2} M^2 (\chi - \chi_i) }{4 \mu^p}
\,.
\ee
Eq.~(\ref{dxi:dN:ph1})
gives the number of e-folds elapsing from the point $(\xi_i,\chi_i)$
until reaching $(\xi_{2i},\chi_2)$, where phase~1 ends,
\be
N_1=N_2^t-N^t_i = - \frac{\mu^p (\xi_{2i} - \xi_i)}{p \Mpl^2 \phic^{p-2}}
\,,
\ee
where $\xi_{2i} =-\sqrt{\xi_i^2 + p \phic^{p-2} M^2 (\chi_2 -  \chi_i ) / (4 \mu^p)}$,
$N_i^t$ is the value of the parameter $N^t$ at the initial point
$(\xi_i,\chi_i)$,
$N_2^t$ its value at $(\xi_{2i},\chi_2)$,
and $\chi_2$ is again given by Eq.~(\ref{chi:2}).

The number of e-folds between the point $(\xi_i,\chi_i)$ and the
point where inflation ends in phase~2 due to the violation of the
slow-roll conditions is then given by $N_1+N_2$, where
$N_2$ are the e-folds between the onset of phase~2 $(\xi_{2i},\chi_2)$
and the violation of slow roll.
For the following purposes, we can again drop the index $i$ on the initial field
values, i.e.  $(\xi_i,\chi_i)\to(\xi,\chi)$ and $(\phi_i,\psi_i)\to(\phi,\psi)$.
The e-fold first derivatives then read
\be \label{eq:dNdphi_ph1}
N_{,\phi} = \frac{1}{\phi_k} \left( N_{1,\xi} + N_{2,\xi}  \right)
\,,
\ee
\be \label{eq:dNdpsi_ph1}
N_{,\psi} = \frac{1}{\psi_k} \left( N_{1,\chi} + N_{2,\chi}   \right)
\,.
\ee
We first evaluate Eq.~(\ref{eq:dNdpsi_ph1}).  The first term gives
\be
N_{1,\chi} = -\frac{ \mu^p }{p \phic^{p-2} \Mpl^2}
 \frac{\dd \xi_{2i}}{\dd \chi} = \frac{M^2}{8 \Mpl^2 \xi_{2} }
\,,
\ee
and the second term yields
\be
N_{2,\chi}   
= - \frac{  p \phic^{p-2} M^2}{16 \xi_2 \mu^p} \frac{\dd N_2}{\dd \xi_{2i} }
\,.
\ee
Using Eq.~(\ref{eq:Ninphase2}) with $\xi_{2i} $ and $\chi_{2i}$ instead of $ \xi_i$ and $\chi_i$ for the number of e-folds realized in phase~2, 
one gets
\be
\frac{\dd N_2}{\dd \xi_{2i} }  = \frac{M^2}{8 \Mpl^2 \xi_{2i}^2} \left( 1 - \frac 2 3 + \frac{2 \xi_{2i}^3}{3 \xiend^3} \right) \simeq \frac{M^2}{24 \Mpl^2 \xi_{2i}^2}
\,,
\ee
and thus one obtains
\be
N_{2,\chi} =
 - \frac{  p \phic^{p-2} M^4}{384 \xi_2^3 \Mpl^2 \mu^p} = - \frac{1}{12 \chi_2} N_{1,\chi}\,,
\ee
where the last expression is found by inserting the value of $\xi_2$
from Eq.~(\ref{xi2:chi2}). Since $\chi_2 > \mathcal O (1)$ [it is a large logarithm {\it cf.} Eq.~(\ref{chi:2})], the dominant term 
in Eq.~(\ref{eq:dNdpsi_ph1}) comes from the number of e-folds realised during phase~1.  In the following, we will therefore neglect the term $N_{2,\chi}$, such that
\be
N_{,\psi} \simeq \frac{M^2}{ 8 \Mpl^2 \xi_{2} \psi}\,.
\ee
We now turn to Eq.~(\ref{eq:dNdphi_ph1}).  For the first term, one finds
\be
\label{dN1dxi}
\frac{\dd N_1}{\dd \xi} 
=  - \frac{\mu^p}{p \Mpl^2 \phic^{p-2} } \left( \frac{\xi}{\xi_2}-1 \right)
\ee
and for the second term,
\be
\frac{\dd N_2}{\dd \xi}
= \frac{\dd N_2}{\dd \xi_{2i}} \frac{\xi}{\xi_2} \approx \frac{M^2 \xi}{24 \Mpl^2 \xi_2^3}
\,.
\ee
Using the appropriate
expression from Eq.~(\ref{xi2:chi2}) for $\xi_2 ^2$, one sees that
\be
\frac{\dd N_1}{\dd \xi} = 6 \chi_2 \left( \frac{\xi_2}{\xi} -1 \right) \frac{\dd N_2}{\dd \xi}
\,.
\ee
Hence again, we may neglect the contribution coming from the number of e-folds in phase~2 and obtain from Eq.~(\ref{dN1dxi})
\be
 N_{,\phi} \simeq \frac{\mu^p}{p \Mpl^2 \phic^{p-1}}~.
\ee
Moreover,
since $\psi_* \ll \psi_2$, we may use that
$ N_{,\phi} \ll N_{,\psi}$.

For the second derivatives, one can again show that the dominant terms come
from the number of e-folds in phase~1.
One therefore obtains
\ba
N_{,\psi \psi}
& = & - \frac{M^2}{8 \Mpl^2 \xi_2 \psi^2} \left( 1 + \frac{1}{2 \chi_2}   \right) \notag\\
& \simeq & - \frac{M^2}{8 \Mpl^2 \xi_2 \psi^2}  \simeq  - \frac{1}{\psi} N_{,\psi}
\,,
\ea
\ba
N_{,\phi \psi}
& \simeq & - \frac{M^2 \xi}{8 \Mpl^2 \xi_2^3 \phic \psi} \simeq  -  \frac{\xi}{\xi_2^2 \phic} N_{,\psi}
\,,
\ea
and
\ba
N_{,\phi \phi}
& = & \frac{\mu^p}{ p \Mpl^2 \phic^{p-2} \phi^2  } \left( -1 + \frac{\xi}{\xi_2} - \frac{1}{\xi_2} + \frac{\xi^2}{\xi_2^3} \right) \\
& \simeq & - \frac{\mu^p}{ p\Mpl^2 \phic^{p-2} \xi_2 \phic^2  }
\,.
\ea

Now we can evaluate the parameter $\fNL$ in the limit where the derivatives with respect to the field $\phi$ as well as the contribution from the number of efolds in phase 2 are negligible.   One finds
\ba
\fNL & \simeq & - \frac 5 6 \frac{N_{,\psi \psi} (N_{,\psi})^2}{ \left( N_{,\psi} \right)^4}
\simeq \frac{20 \Mpl^2 \xi_2 }{3 M^2} \notag\\
  & \simeq & - \frac{10\sqrt p \Mpl^2 \phic^{p/2-1}}{3 M \mu^{p/2}}\sqrt{\chi_2}
\,. \label{eq:fNLphase1}
\ea
For the F-term model, this implies
\be
|\fNL | \simeq  \frac{5 \kappa \sqrt{\cal N}\sqrt{\log 2} \Mpl^2}{6 \sqrt 2 \pi m^2 } \sqrt \chi_2\lesssim 0.13 \sqrt{\chi_2}
\ee
where the maximal negative value of $\fNL$ is obtained when 
the waterfall lasts just about 60 e-folds, i.e. when $\kappa \sqrt{{\cal N}} \approx m^2 / \Mpl^2$ \cite{Clesse:2012dw}.  Particle physics experiments impose a lower bound on $m$, inducing the upper bound $\sqrt{\chi_2} \lesssim 6 $.  The negative amplitude of the $\fNL$ parameter is therefore never higher than about unity, which is below the Planck sensitivity.  We have plotted in FIG.~\ref{fig:fNL} the level of local non-gaussianities given by Eqs.~(\ref{eq:fNLphase1}) and (\ref{eq:fNLphase2}), as a function of $\kappa$ for the F-term model with different values of $m$. 
For the D-term model, the level of non-gaussianities is given by the same expression, with $m_{\rr{FI}}$ instead of $m$.  It is independent of the model parameter $g$, apart
logarithmically through the $\sqrt{\chi_2}$ factor.

For the original model, requiring 60 e-folds along the waterfall imposes $M \mu \gtrsim 40 \Mpl^2$ ~\cite{Clesse:2010iz}.  
The level of non-gaussianities is independent of the parameters $\Lambda$ and $\phic$ (apart through $\sqrt{\chi_2}$),
and its maximal negative value
is therefore of about $\fNL \simeq - 0.3$, as for the F-term and D-term models.  In TABLE~\ref{tab:original}  we give the level of non-gaussianities from the analytical approximations
in the original model
for various parameter sets, covering the qualitatively
different regimes, and compare to the numerical results. 
Notice finally that the particular case where the temporal 
minimum is reached in phase~1 is not relevant, because
that would imply that $\chi_2 \ll 1$ and thus the quantum diffusion would still be
dominating at the time of Hubble exit of the observable scales. 

At this point, we can justify our choice of a final hypersurface fixed at $\xi = \xiend$, instead of a surface of constant energy density.  Considering only the phase-2, one gets that $\psi_{\rr{end}}$ is given by 
\be
\psi_{\rr{end}}^2 = 2 \phic^2 \left( \xiend^2 - \xi_{\rr i}^2 \right) + \psi_{\rr i}^2
\ee
In the context of the $\delta N$ formalism, one can evaluate the shift $\Delta (\psi_{\rr{end}}^2)$ induced by a perturbations of the fields at the time of Hubble crossing of the pivot scale.  For a perturbation of the auxiliary field, one gets
\be
\frac{\Delta (\psi_{\rr{end}}^2)}{\Delta \psi_*} = 2 \psi_*  = 2 \sqrt 2 \phic \xi_*
\ee
whereas a perturbation of the $\phi$ field leads to
\be
\frac{\Delta (\psi_{\rr{end}}^2)}{\Delta \phi_*} = 4 \phic \xi_*
\ee
The potential at the end of inflation is given by
\be
V(\phi_{\rr{end}}, \psi_{\rr{end}}) = \Lambda \left( 1 + 2 \xiend \frac{ \psi_{\rr{end}}^2}{M^2}  \right)
\ee
and therefore, in order to reach a surface of constant density, the shift in $\psi_{\rr{end}}$ should be compensated by a shift in $\xi_{\rr{end}}$, which reads respectively
\be
\frac{\Delta \xiend }{\Delta \psi_*} = \frac{ 4 \sqrt 2 \phic \xi_*}{M^2}~,
\ee
\be
\frac{\Delta \xiend }{\Delta \phi_*} = \frac{ 8  \phic \xi_*}{M^2}~.
\ee
Considering that $\xi_* \simeq  - M^2 /(8 \Mpl^2 N_*) $, one obtains that 
\be
\frac{\Delta \xiend }{\Delta \psi_*} = \frac{ \sqrt 2}{2} \frac{\Delta \xiend }{\Delta \phi_*}  \ll 1~.
\ee  
In the limit $| \xi_* | \gg | \xiend |$, the corresponding shift in Eqs. (\ref{eq:Nphi_phase2}), (\ref{eq:Npsi_phase2}), (\ref{eq:Nphiphi_phase2}), (\ref{eq:Npsipsi_phase2}) and (\ref{eq:Nphipsi_phase2}) induced by the different choices for the final hypersurface can therefore be safely neglected.  

In the case where the pivot scale exits the Hubble radius in phase-1 one can evaluate the shift of $\xi_2^2$ as 
$\Delta \xi_2^2 = - \Delta \xi_*^2 $ and $\Delta \xi_2^2 = - p \phic^{p-2} M^2 \Delta \chi_* /(4 \mu^p) $ for field perturbations respectively in the longitudinal and transverse directions.  By using $\Delta (\psi_{\rr{end}}^2) = - 2 \phic^2 \Delta \xi_2 ^2$ and then
$\Delta \xiend = 2 \Delta \psi_{\rr{end}}^2 / M^2 $, it is straightforward to show that 
\be
\frac{\Delta \xiend }{\Delta \psi_*} \ll 1, \hspace{1cm} \frac{\Delta \xiend }{\Delta \phi_*}  \ll 1~.
\ee  
Since the dominant terms in $N_{,\phi}$, $N_{,\psi}$, $N_{,\phi \phi}$, $N_{,\psi \psi}$ and $N_{,\phi \psi}$ come from the variation of the number of e-folds in phase 1, and because here again the shift in the terms coming from the number of e-folds in phase 2 can be neglected, one can conclude that our results are independent of the possible choices for the final hypersurface.

\subsection{Numerical analysis}\label{subsec:numeric}

The analytical results of the previous Section are valid under some approximations, namely: 
i) the slow-roll approximation, ii) a sharp transition between phase~1 and phase~2, iii) the final hypersurface given by the condition $\xi = \xiend$ [according to Eq.~(\ref{xi:end})] instead of a uniform density condition, iv) the Taylor expansion~(\ref{eq:Ninphase2}) of the number of e-folds in phase~2,
v) some terms neglected in the e-folds derivatives.  To check the validity and the accuracy of our results, 
we have implemented a numerical calculation that
makes use of the $\delta N$ formalism.  Practically, we use the following algorithm:
\begin{enumerate}
\item First, a numerical integration of the exact multi-field background dynamics, from the critical instability point, where we set $N^t=0$, until the end of inflation in order to obtain the total number of e-folds $N^t_{\rm end}$ of inflation along that part of the field trajectory.
\item Second, an integration of the background dynamics from the critical point down to the time $t_*$ of Hubble exit of the pivot scale $k_*=0.05 \rr{Mpc}^{-1}$. With the help
of the first step, we know that the exit point is reached when the e-fold parameter
equals $N^t_*=N^t_{\rm end}-N_{\rm exit}$.
\item Numerical integration of the field dynamics from initial conditions on a $3\times 3$ grid of values centered on $(\phi_*, \psi_*)$. Determination of the number of e-folds $N$
to reach the final hypersurface.
\item Numerical evaluation of the derivatives $\Ni$ and $\Nij$.  The $\fNL$ parameter as well as the amplitude $\mathcal P_{\zeta}$ and the spectral tilt $\ns$ of the power spectrum of curvature perturbations can then be computed by using Eqs.~(\ref{eq:deltaN_fNL}),~(\ref{eq:deltaN_As2}) and~(\ref{eq:deltaN_ns}).
\end{enumerate}
For the stability of the code, notice that the differences between the initial conditions need to be carefully chosen,
sufficiently small for the numerical derivatives $\Ni$ and $\Nij$ to be accurate, but at the
same time sufficiently large
for the differences between the values of $N$ to be much larger than the integration steps 
(that cannot be lower than $\Delta N \sim 10^{-4}$ without increasing unreasonably the integration computing time).

Another numerical issue is related to the choice of the final hypersurface.  It is particularly tricky to define it numerically because the variation of the false vacuum potential along the waterfall trajectories is so tiny that it cannot be resolved due to the limited numerical precision.  We have thus considered the following alternatives: i) by defining the final hypersurface as the end of the slow-roll regime, more precisely when one of the slow-roll parameters reaches unity, ii) by implementing the uniform energy condition after the end of inflation, when for instance $\rho_{\rr{end}} = 0.99 \rho_{\rr{inf}}$ (in this case the effects of the tachyonic preheating are considered to be negligible and the classical trajectories are assumed to be valid down to the final hypersurface), iii) by defining a hypersurface of uniform potential energy density without including the constant term $V_0 = \Lambda$.  We did not find noticeable differences between these possible methods.

At the critical instability point, the classical dynamics of the auxiliary field is dominated by the quantum diffusion.  Classical trajectories must therefore be seen as emerging from this quantum stochastic regime. They are valid once the classical vacuum expectation value
of the auxiliary field fulfills
$\psi \gg \sqrt{\langle \psi_{\rr{qu}}^2 \rangle}$, where $\psi_{\rr{qu}}$ is
the operator for the quantum field fluctuations around the classical expectation value $\psi$.  For the numerical integration of the field dynamics, we take for the initial condition
of the auxiliary field at the critical instability point the value
given by the process of quantum diffusion~\cite{Clesse:2010iz,Clesse:2012dw}
\ba
\psi_0 & \simeq & \frac{\sqrt{\kappa} M_{\rr{F}}^3}{2 \sqrt 3 \pi^{3/4} (\ln 2)^{1/4}} \  \mbox{ for the F-term}\,, \label{eq:psi0Fterm}\\
\psi_0 & \simeq & \frac{g m_{\rr{FI}}^3}{8  \sqrt{3 \kappa} \pi^{3/4} (\ln 2)^{1/4}} \  \mbox{ for the D-term}\,, \label{eq:psi0Dterm} \\
\psi_0 & \simeq & \sqrt{ \frac{\Lambda \mu M  }{96  \pi^{3/2}}}  \ \mbox{ for the original model} \label{eq:psi0original}
\ea
and only consider the trajectory as valid when $\psi \gg \psi_0$.  

\begin{figure}[!ht]
\begin{center} 
\includegraphics[height=55mm]{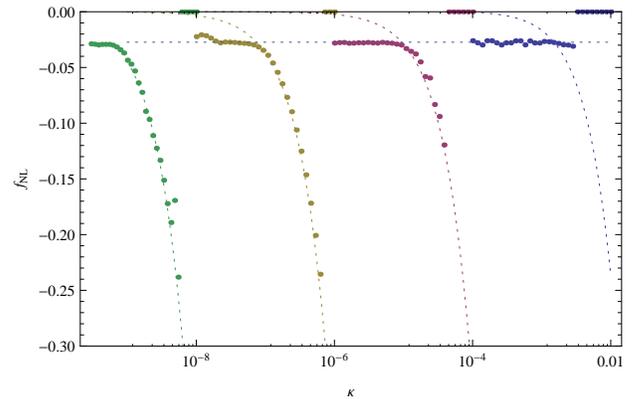}
\caption{\label{fig:fNL}
Local $\fNL$ parameter as a function of $\kappa$, for the F-term model, with from left to right
$m = 10^{-4} \Mpl$, $m = 10^{-3} \Mpl$, $m = 10^{-2} \Mpl$, $m = 10^{-1} \Mpl$.  
The bold dots are the numerical results using the $\delta N$ formalism.  The dotted horizontal line corresponds to $\fNL = -5 /[3(\Nexit +1) ] $ [the approximate result from Eq.~(\ref{eq:fNLphase1})]
and the dashed lines to Eq.~(\ref{eq:fNLphase1}), which are respectively valid for $t_* > t_{1,2}$ and $t_* < t_{1,2}$.  We assume for simplicity that $\Nexit = 60$.
}
\end{center}
\end{figure}

For the F-term model, we have plotted in Fig.~\ref{fig:fNL} the local $\fNL$ parameter calculated numerically as a function $\kappa$, for different values of the mass parameter $m$ in the range $10^{-4} \Mpl < M < 0.1 \Mpl$.  For comparison, the analytical approximations derived in the previous Section are also presented.  Numerical and analytical results agree well, with $\fNL$ values ranging typically from $-0.03$ down to $-0.3$,
as long as $m \lesssim 0.01 \Mpl$.  For larger values, in the regime $t_* < t_{1,2}$, our approximation given by Eq.~(\ref{eq:fNLphase1}) is not valid any more and the $|\fNL|$ parameter takes lower values.  This is because the e-folds derivatives with respect to the field $\phi$ become an important contribution to the level of non-gaussianities.  We can draw identical conclusions for the D-term model, where the level of non-gaussianitiy is nearly independent of the coupling $g$. 

For the original model, the parameter space has two additional dimensions and it is more difficult to explore it entirely with our numerical method.  However, we present in TABLE~\ref{tab:original} a comparison of numerical and analytical results for a few sets of parameters corresponding to the different regimes. This serves as a check that the numerical results are in agreement with the analytical approximations.  


\begin{figure}[!ht]
\begin{center}
\includegraphics[height=55mm]{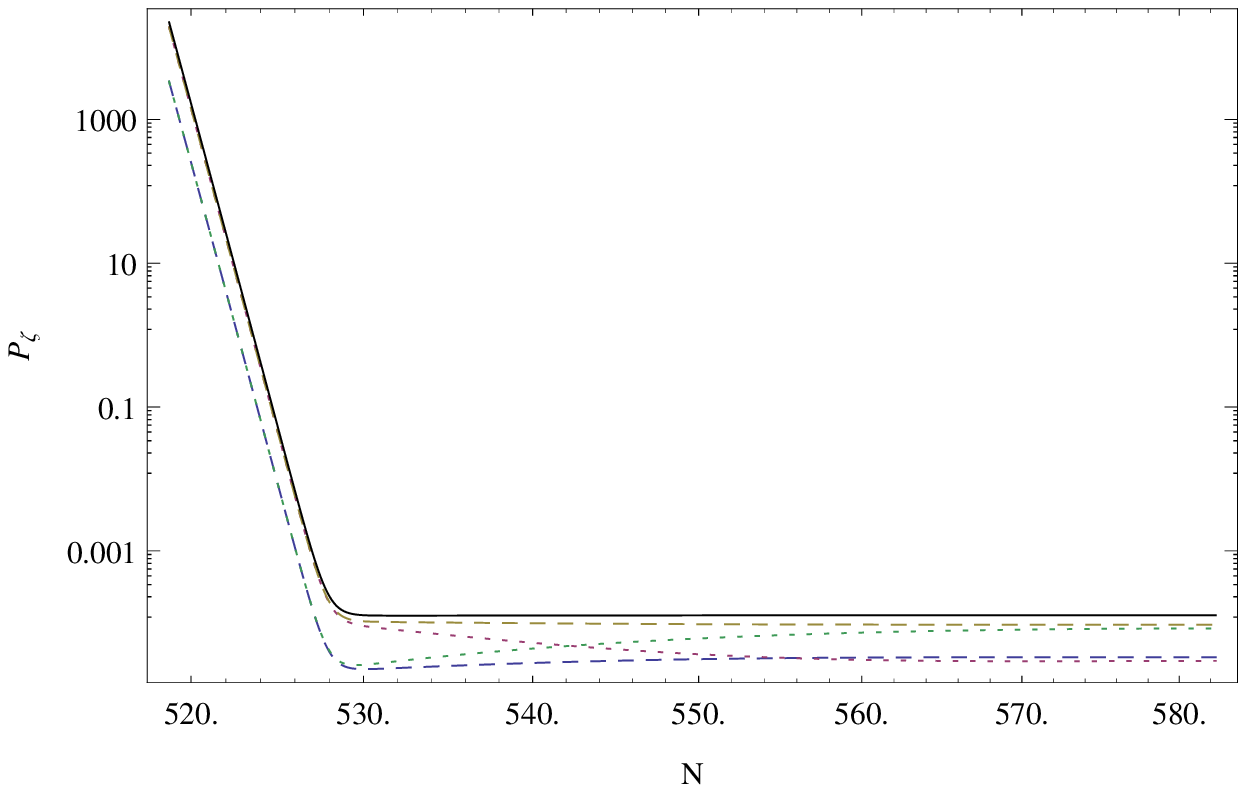} 
\caption{ \label{fig:evolpert_largeN}
Evolution of the power spectrum of curvature perturbations (black) as well as of rescaled adiabatic (dashed) and entropic (dotted) perturbations, generated respectively by
initial perturbations of $\phi$ (blue and red curves) and $\psi$ (green and yellow curves), and the power spectrum of curvature perturbations (black curve) for a pivot scale $k_* = 0.05 \rr{Mpc}^{-1}$ and F-term potential parameters $m = 10^{-3}$ and $\kappa = 5 \times 10 ^{-8}$. These parameters correspond to the case $t_* > t_{1,2}$.
}
\end{center}

\begin{center}
\includegraphics[height=55mm]{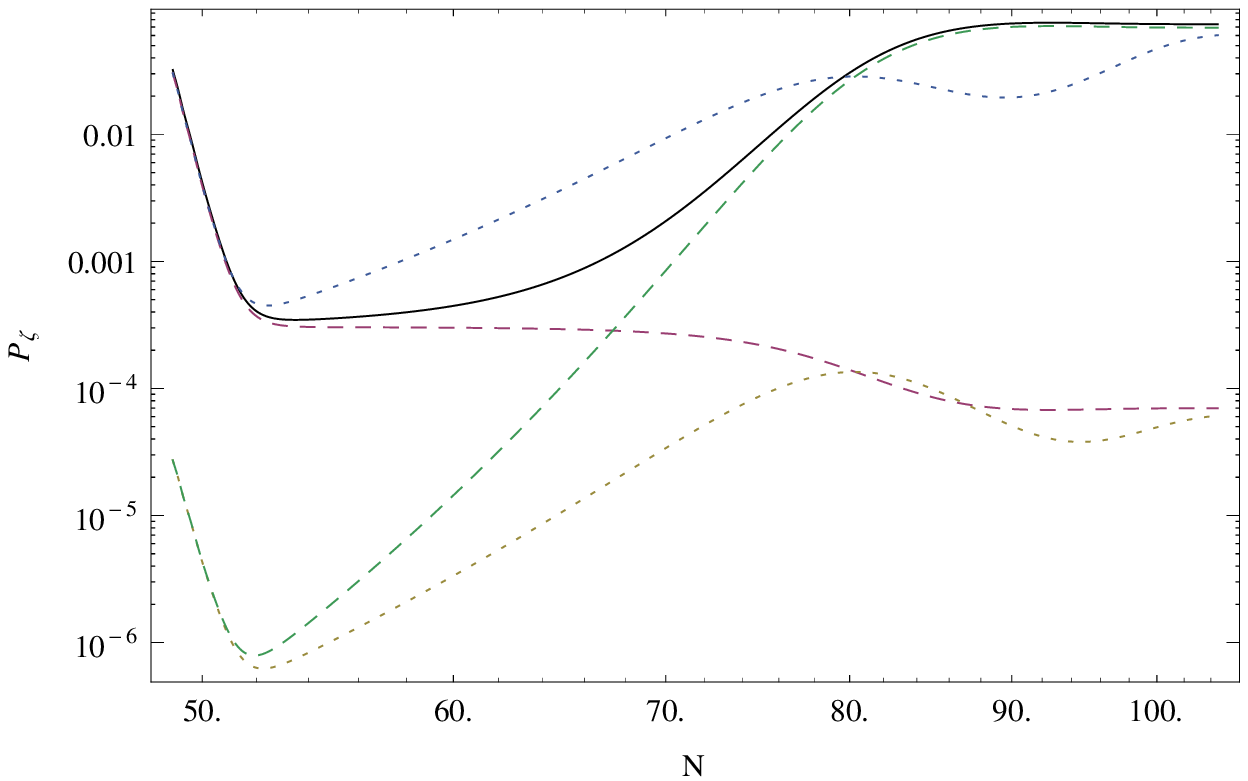}
\caption{\label{fig:evolpert_smallN}
Evolution of the power spectrum of curvature perturbations (black) as well as of rescaled adiabatic (dashed) and entropic (dotted) perturbations, sourced respectively by
initial perturbations of $\phi$ (blue and red curves) and $\psi$ (green and yellow curves), and the power spectrum of curvature perturbations (black curve) for a pivot scale $k_* = 0.05 \rr{Mpc}^{-1}$ and F-term potential parameters $m = 10^{-3}$ and $\kappa = 3 \times 10 ^{-7}$.
These parameters correspond to the case $t_* < t_{1,2}$}
\end{center}
\end{figure}


\begin{table*}[!ht]
\begin{tabular}{|c|c|c|c|c|c|c|c|}
\hline 
Parameters & Regime & $\fNL ^{\rr{num}}$ & $\fNL ^{\rr{app}} $ &  $\ns ^{\rr{num}}$ & $\ns ^{\rr{app}}$ &  $ {\cal P}_\zeta^{\rr{num}} (k_*) $ 
& $ {\cal P}_\zeta^{\rr{app}} (k_*) $  \\
\hline
$M = \phic = 0.01 \ \Mpl, \mu= 10^5 \  \Mpl, \Lambda = 10^{-20} \ \Mpl^4$ & $t_{1,2} < t_*$ & $-0.030$ & $-0.029$ 
& $0.929$ & $0.930$ &  $0.017$ & $0.019$ \\
$M = \phic = 0.001 \ \Mpl, \mu= 10^6 \  \Mpl, \Lambda = 10^{-30} \ \Mpl^4$ & $t_{1,2} < t_*$ & $-0.033$ & $-0.033$
& $0.921$ & $0.921 $ &  $1.1 \times 10^{-6}$ & $1.3 \times 10^{-6}$ \\
$M = 0.001 \ \Mpl, \phic=10^{-5}, \mu= 10^6 \  \Mpl, \Lambda = 10^{-30} \ \Mpl^4$ & $t_{1,2} < t_*$ & $-0.033$ & $-0.032$
& $0.921$ & $0.921 $ &  $0.0011 $ & $0.012$ \\
$M = 0.01 \ \Mpl, \phic = 0.1 \ \Mpl, \mu= 10^5 \  \Mpl, \Lambda = 10^{-20} \ \Mpl^4$ & $t_{1,2} < t_* < t_{TM} $ 
& $-0.030$ & $-0.030$  & $0.929$ & $0.929$ &  $1.8 \times 10^{-4}$ & $1.9 \times 10^{-4}$  \\
$M = 0.001 \ \Mpl, \phic = 0.1 \ \Mpl, \mu= 10^6 \  \Mpl, \Lambda = 10^{-30} \ \Mpl^4$ & $t_{1,2} < t_* < t_{TM} $ 
& $-0.033$ & $-0.032$ & $0.921$ & $0.921$ &  $1.1 \times 10^{-10}$ & $1.3 \times 10^{-10}$ \\
$M = \phic = 0.01 \ \Mpl, \mu= 10^4 \  \Mpl, \Lambda = 10^{-20} \ \Mpl^4$ & $t_* < t_{1,2}$ & $-0.10$ & $-0.11$ 
& $0.881$ & $0.880$ &  $0.23$ & $0.29$ \\
$M = \phic = 0.01 \ \Mpl, \mu= 10^{3.8} \  \Mpl, \Lambda = 10^{-20} \ \Mpl^4$ & $t_* < t_{1,2}$ & $-0.17$ & $-0.18$ 
& $0.946$ & $0.955$ &  $1.2$ & $1.36$ \\
$M = \phic = 0.001 \ \Mpl, \mu= 10^5 \  \Mpl, \Lambda = 10^{-30} \ \Mpl^4$ & $t_* < t_{1,2}$ & $-0.14$ & $-0.15$
& $0.78$ & $0.77$  & $2.9 \times 10^{-4}$ & $4.6 \times 10^{-4}$  \\
$M = \phic = 0.001 \ \Mpl, \mu= 10^{4.8} \  \Mpl, \Lambda = 10^{-30} \ \Mpl^4$ & $t_* < t_{1,2}$ & $-0.23$ & $-0.24$ 
& $0.77$ & $0.77$  & $0.021$ & $0.035$  \\
$M = 0.001 \ \Mpl, \phic=10^{-4} \Mpl, \mu= 10^5  \Mpl, \Lambda = 10^{-30} \Mpl^4$ & $t_* < t_{1,2}$ & $-0.12$ & $-0.13$
& $0.82$ & $0.81$  & $6.1 \times 10^{-3}$ & $8.4 \times 10^{-3}$  \\
$M = 0.001 \ \Mpl, \phic=10^{-4} \Mpl, \mu= 10^{4.8}  \Mpl, \Lambda = 10^{-30} \Mpl^4$ & $t_* < t_{1,2}$ & $-0.20$ & $-0.21$ 
& $0.83$ & $0.84$  & $0.15$ & $ 0.21 $  \\
\hline
\end{tabular}
\caption{\label{tab:original}
Comparison of the power spectrum, its tilt and the non-gaussianities in the original
model for various illustrative points in parameter space, representing the horizon exit of
the pivot scale during the qualitatively different phases. We compare the approximate analytical
results with the numerical results obtained using the $\delta N$ formalism, assuming instantaneous reheating.
}
\end{table*}

\section{Power spectrum of curvature perturbations and contribution of entropic modes} \label{sec:power_spectrum}

\subsection{Using the $\delta N$ formalism}

The amplitude and spectral index of the power spectrum of curvature perturbations at the end of inflation can be calculated by using the $\delta N$ formalism, as explained in Section~\ref{sec:deltaN}.  For the spectral index, in addition to the e-folds derivatives that have been calculated in the previous Section, 
one needs to know the field derivatives at the time of Hubble exit of the pivot scale.

\subsubsection{Case 1:  Hubble exit in phase 2}

In the generic case where $t_* > t_{1,2}$,  one obtains
\ba
\frac{\dd \phi}{\dd N^t} &=& -\frac{8 \Mpl^2 \phic \xi^2}{M^2} = -\frac{1}{3 N_{,\phi}}\,,
\\
\frac{\dd \psi}{\dd N^t} &=& \sqrt 2 \frac{8 \Mpl^2 \phic \xi^2}{M^2}  = -\frac{1}{3 N_{,\psi}}\,.
\ea
Because observable scales exit the Hubble radius near the critical instablitiy point, one has $ \epsilon_* \ll 1$, 
so that it can be neglected.  Using Eq.~(\ref{eq:deltaN_ns}), one then finds
\ba
\ns - 1 
 & \simeq & \frac{32 \xi_* \Mpl^2}{M^2}  \simeq  - \frac{4}{ \left(\Nexit - \frac{M^2}{8 \Mpl^2 \xiend}  \right)}
\,.
\ea
This formula corresponds to the one derived in Refs.~\cite{Kodama:2011vs,Clesse:2012dw}
assuming that the waterfall trajectories are effectively single-field.
For the amplitude of the power spectrum of curvature perturbations, one gets
\ba 
\label{P:phase2}
{\cal P}_\zeta(k_*)  
 & \simeq & \frac{\Lambda M^4}{16 \times 24^2 \pi^2 \Mpl^6 \phic^2 \xi_*^4 } \notag\\
 & \simeq & \frac{4 \Lambda \Mpl^2 (\Nexit -  \frac{M^2}{8 \Mpl^2 \xiend})^4 }{9 \pi^2 M^2 \phic^2  }
\label{eq:ns_case1}
\ea
Here also the results of Refs.~\cite{Kodama:2011vs,Clesse:2012dw} are recovered. 
Therefore the regime corresponding to $t_* > t_{1,2}$ is effectively single field.  
The F-term and D-term models generate a red tilted power spectrum, but the spectral index is too low when compared to CMB observations.  
For the original model, it is in principle possible to increase its value up to $\ns \simeq 0.94$ 
by increasing $\Lambda$ that is a free parameter, and therefore $\Nexit$. However, in this
case inflation is realised at an energy scale near the limit imposed by the
search for $B$-mode polarization in the CMB.

Finally, for the original model, notice that our comment regarding the specific case where the trajectories reach the temporal minimum before the end of inflation still applies, and $\xiend$ can be replaced by $\xi_{2.\rr{TM}}$ in Eq.~(\ref{eq:ns_case1}).

\subsubsection{Case 2:  Hubble exit in phase 1}

\begin{figure*}[h]
\begin{center}
\includegraphics[height=100mm]{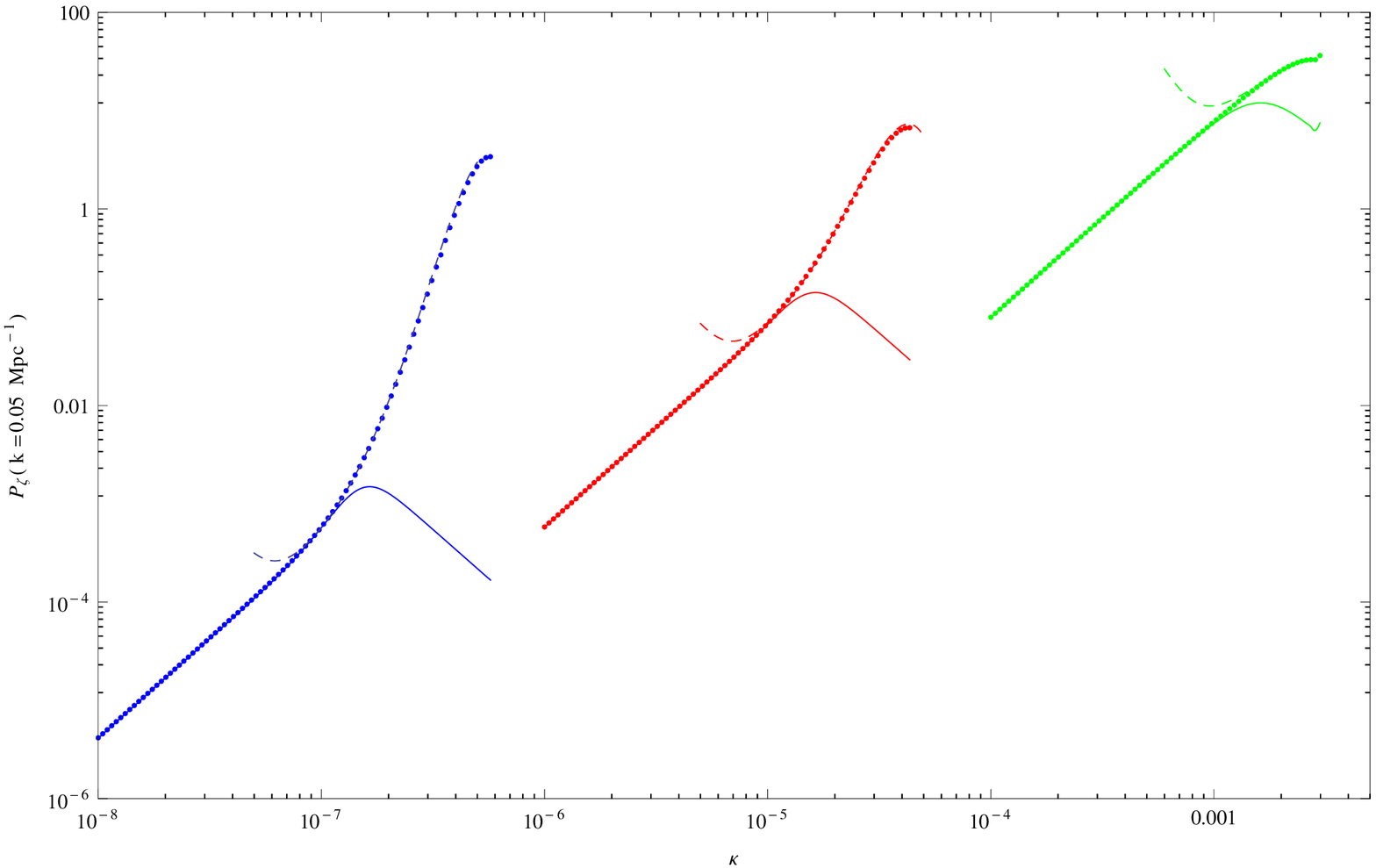}
\caption{ \label{fig:Pk}
Amplitude of the power spectrum of curvature perturbations from the
analytic approximations based on the $\delta N$ formalism (Hubble exit in phase 1: dashed),
from the numerical integration (points), and assuming effectively single field trajectories
(solid) as a function of the $\kappa$ parameter for the F-term model, for a pivot scale $k_* = 0.05 {\rm Mpc}^{-1}$, and from left to right, $m = 10^{-3} / 10^{-2} / 10^{-1}$.
}
\end{center}

\begin{center}
\includegraphics[height=100mm]{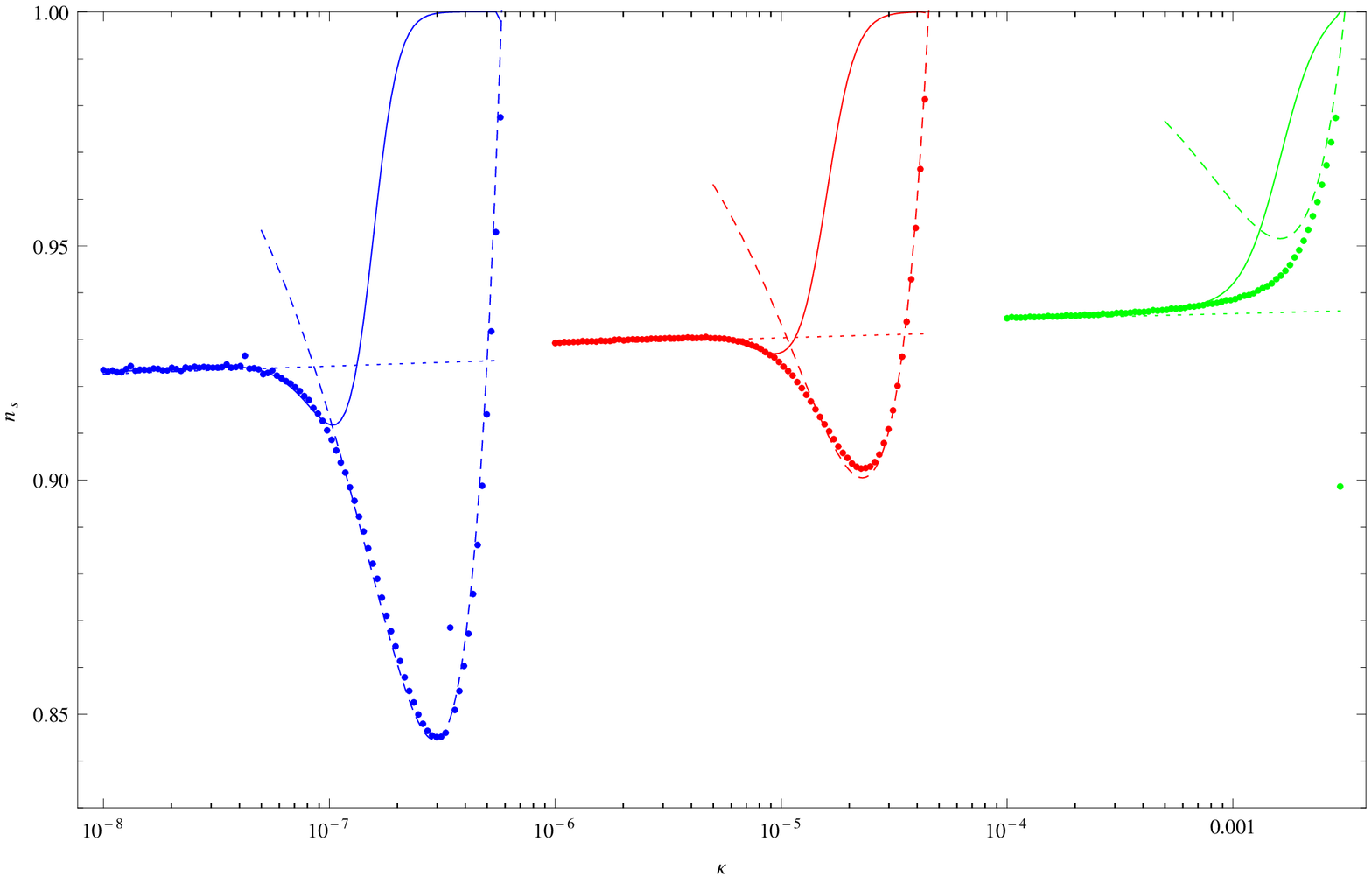}
\caption{ \label{fig:ns}
Spectral index of the power spectrum of curvature perturbations
derived from the
analytic approximations based on the $\delta N$ formalism (Hubble exit in phase 1: dashed / phase 2: short dashed),
derived from the numerical integration (points), and assuming effectively single field trajectories (solid), 
as a function of the $\kappa$ parameter for the F-term model, for a pivot scale $k_* = 0.05 {\rm Mpc}^{-1}$, and from left to right, $m = 10^{-3} / 10^{-2} / 10^{-1}$.  
}
\end{center}
\end{figure*}

In the case where the potential parameters are tuned such that $t_* < t_{1,2}$, we find
\ba
\frac{\dd \phi}{\dd N^t} &=& - \frac{p \phic^{p-1} \Mpl^2}{\mu^p} = -\frac{1}{ N_{,\phi}}~,
\\
\frac{\dd \psi}{\dd N^t} &=& - \frac{8 \Mpl^2 \psi \xi}{M^2}  = -\frac{\xi}{N_{,\psi} \xi_2}~.
\ea
By using these equations as well as the relations $N_{,\psi} \gg N_{,\phi}$, $N_{,\psi \psi} \gg N_{,\psi \phi} $ 
and $N_{,\psi \psi} \gg N_{,\phi \phi}$, we obtain the leading term for the spectral index
\be
\label{eq:ns_case2}
\ns - 1 \simeq \frac{16 \Mpl^2 \xi_*}{M^2} 
\,.
\ee
When $\xi_* \rightarrow \xi_2$ it is connected to the spectral index for $t_* > t_{1,2}$.  
Notice that instead of increasing when the Hubble exit occurs deeper in phase~1, the spectral index first takes lower values than expected for effectively single-field trajectories. This
behaviour is shown in FIG.~\ref{fig:ns} that gives the spectral index as a function of $\kappa$ for the F-term model and different values of $M$.  This is because $|\xi_2 |$, and thus $|\xi_*|$, first increases with $\kappa$.  The spectral index increases up to unity only when $\xi_* \rightarrow 0$.  From Eq.~(\ref{eq:ns_case2}) only, one can conclude that it is possible to find a spectral index value in agreement with CMB observations.  It turns out however, that the power spectrum amplitude is strongly modified in this case by entropy perturbations and cannot fit to CMB observations, as explained in
the following.

The amplitude of the power spectrum is given by
\be \label{eq:Pzeta_exit_in_phase1}
{\cal P}_\zeta (k_*) \simeq \frac{\Lambda M^2 \mu^p}{192 \pi^2 p \Mpl^6 \phic^{p-2} \chi_2 \psi_*^2}
\,.
\ee
In the case where $t_* \simeq t_{1,2}$, we replace
$\psi_*=\psi_0\exp(\chi_2)$, such that the amplitude is
\be
{\cal P}_\zeta (k_*, t_* \simeq t_{1,2}) \simeq \frac{\Lambda \mu^{2p}}{48 \pi^2 p^2 \Mpl^6 \phic^{2p-2} \chi_2 }
\,,
\ee
and it connects continuously to what is found in
Eq.~(\ref{P:phase2}) for
$t_* > t_{1,2} $.  When the Hubble exit
of the scale $k_*$ occurs deeper in phase~1,
we see using Eqs.~(\ref{xi:chi:ph1}) and~(\ref{xi:N:ph1})
that the amplitude grows exponentially as
\be
\label{P:exponential}
{\cal P}_\zeta (k_*) \times \exp \left[ 2 \chi_2 \left( 1 - \frac{{N^t_*}^2}{N_1^2}   \right)   \right]
\,,
\ee
where $N_1$ is the number of e-folds in phase~1 (between the critical point and the transition to phase~2) and $N_t^*$ is the number of e-folds between the critical point
and the
horizon exit of the scale $k_*$ (i.e. the value of the parameter $N^t$ at horizon exit,
provided $N^t=0$ at the critical point).
The spectrum then may
reach a maximal amplitude typically larger than unity,
\be
{\cal P}_\zeta (t_* \simeq 0 ) \simeq \frac{\Lambda M^2 \mu^{p}}{192 \pi^2 p^2 \Mpl^6 \phic^{p-2} \chi_2 \psi_0^2 } 
\,.
\ee
It is important to notice that there is no freedom to fix the amplitude of the curvature power spectrum independently of its spectral index, because $\psi_*$ in Eq.~\ref{eq:Pzeta_exit_in_phase1} is related to $\xi_*$ through Eq.~\ref{xi:chi:ph1} describing the waterfall trajectory in phase-1.  Then, considering that $\psi_0$ takes values given by Eqs.~(\ref{eq:psi0Fterm}),~(\ref{eq:psi0Dterm}) or~(\ref{eq:psi0original}) depending on the model and that $\chi_2$ only depends logarithmically on the model parameters (it is typically of order unity and cannot be used to rescale by a significant amount the amplitude of the spectrum), one can see that for parameter values $\kappa \sim M^2 / \Mpl^2$ or $M \mu \sim \Mpl^2$ (to which corresponds the regime where $\Nend \gtrsim 60$ and Hubble exit of observable modes in phase-1 ), the power spectrum amplitude is several order of magnitudes larger than $10^{-9}$. 

FIGs.~\ref{fig:Pk} and~\ref{fig:ns} for the F-term model illustrate this exponential growth that prevents the amplitude to be in agreement with CMB observations when the spectral tilt is in the allowed range by CMB observations.  As mentioned in the previous Section, the case where the temporal minimum is reached in phase~1 is not relevant because of the quantum diffusion.  

\subsection{From the numerical integration of multi-field perturbations}

We now compare with the results obtained from the methods explained
in Section~\ref{sec:MF_perturbations}.
By integrating the multi-field perturbations, one can follow their sub-Hubble and super-Hubble evolution throughout the waterfall phase and 
 identify the contributions of the fields to the rescaled adiabatic and entropic perturbations, that are respectively defined as~\cite{Ringeval:2007am} 
\ba
\delta \pi_{\rr a} & = & \frac{\dot \phi \delta \phi }{\sqrt{\dot \phi^2 + \dot \psi^2}} + \frac{\dot \psi \delta \psi }{\sqrt{\dot \phi^2 + \dot \psi^2}}\,, \\
\delta \pi_{\rr e} & = & \frac{\dot \psi \delta \phi }{\sqrt{\dot \phi^2 + \dot \psi^2}} + \frac{\dot \phi \delta \psi }{\sqrt{\dot \phi^2 + \dot \psi^2}}
\,.
\ea
We have plotted on FIGs.~\ref{fig:evolpert_largeN} and~\ref{fig:evolpert_smallN} the evolution of curvature, rescaled adiabatic and entropic perturbations for the F-term model (but we find similar behaviors for the D-term and the original models) for the case $t_* > t_{1,2}$ and $t_* < t_{1,2}$, respectively.  

In the first case, the curvature perturbations freeze when they become super-Hubble, as expected for an effectively single field model.  Notice also that in the sub-Hubble regime, 
the adiabatic perturbations induced by the initial conditions for $\delta \phi$ with $\delta \psi$ initially set to zero correspond to the entropy perturbations induced by $ \delta \psi $ with $\delta \phi$ initially set to zero, and conversely.  This is expected because the field perturbations evolve as independent plane waves with identical amplitudes in the sub-Hubble regime, and because 
\ba
\delta \pi_{\rr a} & = & \cos \theta \delta \phi - \sin \theta \delta \psi \,,\\
\delta \pi_{\rr e} & = & \sin \theta \delta \phi + \cos \theta \delta \psi   
\,,
\ea
where $\theta $ is the angle between a unit vector tangential to the field trajectory and the $\phi = 0 $ direction.

In the second case ($t_* < t_{1,2}$), we observe that super-Hubble curvature perturbations are receive contributions
from entropy perturbations that are generated during the field evolution in phase~1 and then are frozen during the phase~2.    
This corresponds to the strong enhancement of the power spectrum amplitude according to Eq.~(\ref{P:exponential})
obtained within the $\delta N$ formalism.

We have compared the analytical approximations of the previous Section to the numerical results for the power spectrum amplitude and spectral index. For the F-term model, we have plotted in Figs.~\ref{fig:Pk} and~\ref{fig:ns} the amplitude and the spectral index of the power spectrum of curvature perturbations as a function of the parameter $\kappa$, for various values of $M$ in the F-term model.  For comparison, we have also plotted the effective 1-field slow-roll predictions and the analytical approximations derived in the previous Section.  These plots illustrate the general agreement we find between the different methods, and the important modification of the power spectrum 
of curvature perturbations when compared to the predictions assuming an effectively single-field dynamics.   
As expected, for the D-term model we find very similar results. 

For the original model, we have compared in TABLE~\ref{tab:original} the power spectrum amplitude and spectral index for some parameter sets corresponding to the different regimes.  
For the spectrum amplitude, we observe a strong discrepancy (up to 30\%) between numerical results and the analytical approximation in the regime where $t_* < t_{1,2}$.  
However, it must be noticed that a tiny modification of $N_2^t - \Nend^t$ can affect importantly the spectrum amplitude in this regime. 
Since this quantity cannot be evaluated analytically with a very good precision (even more since we assume a sharp transition between phase~1 and phase~2, which is not exactly the case), such a strong discrepancy can be explained.  
In the case $t_* > t_{1,2}$, numerical and analytical results agree well.  
Finally, we observe that when the temporal minimum is reached in phase~2 before the end of inflation, the predictions from the numerical and
the approximate analytical method only show marginal deviations, because $|\xi_{2.\rr{TM}}| \approx M^2 /\phic^2 \ll 1$ in this particular regime.

\section{Conclusion}\label{sec:conclusion}

Using two different methods -- the $\delta N$ formalism and the numerical integration of the linear multi-field perturbations -- we have calculated the level of non-gaussianities and the power spectrum of curvature perturbations produced in the parametric
regime of a mild waterfall phase in models of hybrid inflation.

We have investigated the supersymmetric F-term and D-term models as well as the original non-supersymmetric hybrid model.
In order to study these within a unified analysis, we have
introduced a common parametrisation for the different variants
of the potential.  For the F-term and D-term models, the mild waterfall regime occurs in the small coupling limit, when $\kappa \lesssim M^2 /\Mpl^ 2$. For the original model this happens when $\mu M > \Mpl^ 2$.  We have only considered field values lower than Planck mass, as it is commonly imposed for an effective field theory
description to be valid.  The defining feature of the mild waterfall regime is that the last 60 e-folds of inflation are realized after
the fields pass the critical point, such that possible cosmological defects are stretched outside the observable Universe.  The observational predictions are modified, and if one assumes that the waterfall trajectories are effectively single-field, one obtains spectral index values from $\ns = 1 - 4/ \Nexit$, when observable scales leave the Hubble radius in the so-called slow-roll phase~2 where the slope of the potential in the valley direction is dominated by the terms involving the auxiliary field of the waterfall ($t_* > t_{1,2}$), and up to unity when this happens in the first phase ($t_* < t_{1,2}$), where the slope along the inflationary valley dominates.
Prior to the present analysis, the model was therefore considered as possibly in agreement with CMB observations~\cite{Kodama:2011vs,Clesse:2012dw}.  
However, since the scalar field trajectories are turning during
the waterfall regime toward the minima of the potential, one might expect a large contribution of entropic modes to the power spectrum of curvature perturbations, as well as a high level of local non-gaussianities.  The quantification of these contributions and
the resulting signatures and the resolution of the question of the
phenomenological viability of hybrid inflation in the
mild waterfall regime are the main results of the present paper.  
For this purpose, we have derived analytical approximations for the local $\fNL$ parameter and the power spectrum amplitude and tilt, 
which agree well with the numerical results.  

For all the models, we find that the generic regime corresponding to $t_* > t_{1,2}$ is effectively described by single field dynamics, with $\fNL \approx - 5/(3(\Nexit +1)) \approx 0.03 $, which is nearly independent of the potential parameters.  For the spectrum amplitude and the spectral index, we confirm the results of Refs.~\cite{Clesse:2010iz,Kodama:2011vs,Clesse:2012dw} by using both the $\delta N$ formalism and the
linear theory of multi-field perturbations.  The resulting spectral index is outside and below the bounds imposed by CMB experiments, and this regime is therefore strongly disfavored. 

For the particular regime where the parameters are tuned so that $t_* < t_{1,2}$, we find that the waterfall trajectories cannot be considered as effectively single field.  The level of non-gaussianities increases, with negative values of $\fNL$ down to about minus unity. Notice however that in no case we find that the level of non-gaussianity exceeds the recent bounds from the Planck experiments. 
Regarding the power spectrum of curvature perturbations, we find that the entropic modes are an important source of the super-Hubble curvature perturbations, enhancing the power spectrum by several order of magnitudes,
 up to a maximal amplitude larger than unity, which is far from the CMB constraints and which would lead to the formation of
primordial black holes. When scales exit the Hubble radius deeper in the first slow-roll phase, the spectral index first takes lower values than expected for an effectively single field model, and then increases up to unity.  However, values for the spectral
index that are in agreement with CMB constraints always appear in conjunction with a power spectrum amplitude that is much higher than the measured one.  Contrary to what was thought before~\cite{Clesse:2010iz,Kodama:2011vs,Clesse:2012dw}, it is therefore impossible to find parameters in agreement with CMB observations.

We leave for a future work the particular case where the waterfall lasts typically $1< N \lesssim 60$, for which we expect a modification of the slow-roll prediction of inflation along the valley.  We also would like to emphasize that the levels of non-gaussianities are calculated in this paper at the end of the inflationary era.  In principle, they can be subsequently modified during the reheating era.  
In Ref.~\cite{Leung:2012ve}, some models are found for which the non-gaussianities are enhanced during the reheating.  But in the context of a tachyonic preheating phase, applying the $\delta N$ formalism seems to be a very challenging issue.  
 Another interesting perspective would be to forecast the constraints on hybrid models from the CMB distortions generated by the perturbations from the waterfall phase at the end of inflation.  Finally, we would like to mention that in the regime where the dynamics of both the fields is stochastic~\cite{Martin:2011ib}, observable predictions have not been derived so far.  Studying this regime will nevertheless require 
new methods since the classical dynamics is not valid in this case.

\section*{Acknowledgements}
The authors acknowledge
support by the DFG cluster of excellence `Origin and Structure
of the Universe'. In addition, SC has received funding
from the Alexander von Humboldt Foundation,
BG from the Gottfried Wilhelm Leibniz programme of the DFG
and YZ from the Chinese Scholarship council.  The authors warmly thank Christophe Ringeval for useful comments and discussion.

\bibliography{biblio}

\end{document}